\documentclass[prb,aps,twocolumn,showpacs,floats]{revtex4}
\usepackage{graphics}
\usepackage{epsfig}
\usepackage{hyperref}
\usepackage{amsfonts}
\usepackage{amssymb}
\usepackage{amsthm}
\usepackage{newlfont}
\usepackage{amsmath}
\usepackage{url}
\begin{document}

\title{Spectral signatures of the BCS-BEC crossover in the 
excitonic insulator phase of the extended Falicov-Kimball model}
\author{Van-Nham Phan$^{1}$, Klaus W. Becker$^{2}$ and Holger Fehske$^{1}$}

\affiliation{
$^{1}$Institut f{\"u}r Physik, Ernst-Moritz-Arndt-Universit{\"a}t Greifswald, D-17487 Greifswald, Germany\\
$^{2}$Institut f{\"u}r Theoretische Physik, Technische Universit{\"a}t Dresden, D-01062 Dresden, Germany
}


\begin{abstract}
We explore the spontaneous formation of an excitonic insulator 
state at the semimetal-semiconductor transition of mixed-valence
materials in the framework of the spinless Falicov-Kimball model 
with direct $f$-$f$ electron hopping.  Adapting the projector-based 
renormalization method, we obtain a set of renormalization differential 
equations for the extended Falicov-Kimball model parameters 
and finally derive analytical expressions for the order parameter, 
as well as for the renormalized $c$- and $f$-electron dispersions, 
momentum distributions, 
and wave-vector resolved single-particle spectral functions.
Our numerical results proved the valence transition picture,
related to the appearance of the excitonic insulator phase,
in the case of overlapping $c$ and $f$ bands. Thereby the 
photoemission spectra show significant differences between 
the weak-to-intermediate and intermediate-to-strong Coulomb
attraction regimes, indicating a BCS-BEC transition of the 
excitonic condensate. 
\end{abstract}
\date{\today}
\maketitle
\section{Introduction}
\begin{figure}[b]
    \begin{center}
      \includegraphics[angle = -0, width = 0.9\linewidth]{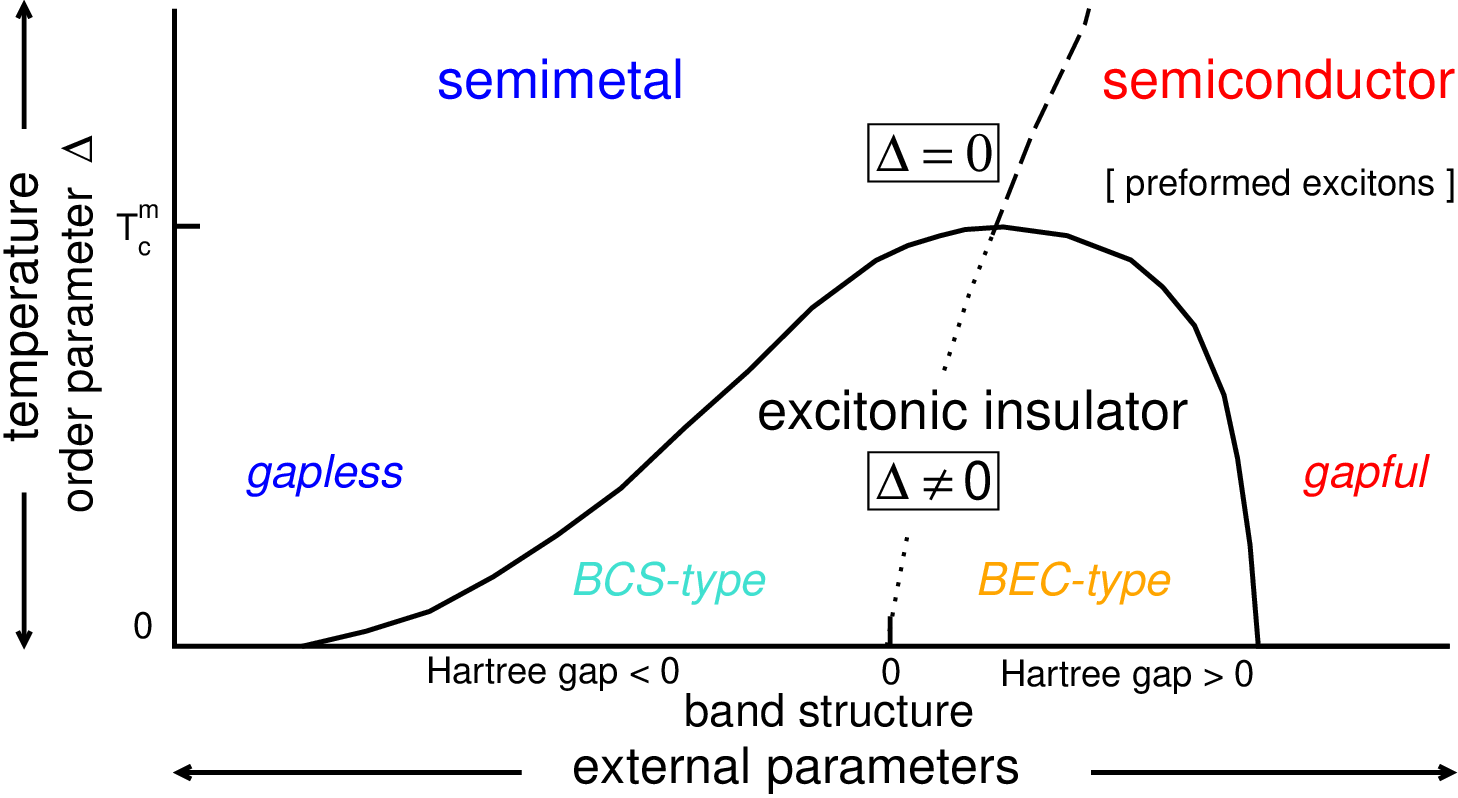}
    \end{center}
\caption{(Color online) 
EI formation and BCS-BEC transition scenario. 
At the semimetal-semiconductor transition the ground state of the 
system may become unstable with respect to the spontaneous formation 
of excitons near the point at which band overlap occurs.  
Starting from a semimetal with small density of electrons and holes  
(such that the Coulomb interaction is basically unscreened), 
the number of free carriers varies discontinuously under an applied 
perturbation, signaling a phase transition~\cite{Mo61}. 
Approaching the transition from the semiconductor side, an anomaly occurs 
when the band gap, tuned, e.g., by external pressure, becomes less 
than the exciton binding energy~\cite{Kno63}. Depending on from 
which side of the semimetal-semiconductor transition the EI is reached, 
the EI can be viewed either as BCS condensate of loosely-bound 
electron-hole pairs or as BEC of preformed tightly-bound 
excitons~\cite{BF06}. A finite order parameter $\Delta$
indicates the new distorted phase of the crystal, with coherence between
conduction- and valence-band electrons and a gap for charge excitations.}
\label{fig:PD}
\end{figure}

The idea that an excitonic phase appears---under certain
circumstances---at the semiconductor-semimetal transition
dates back about half a century~\cite{Mo61,Kno63}.
The formation of excitons is driven by the Coulomb 
attraction between conduction-band electrons and valence-band holes. 
Provided a large enough number of sufficiently  
long-lived excitons was created, a subsequent
spontaneous condensation of these composite Bose 
quasiparticles may set in. 
The excitonic instability is expected to happen,   
when semimetals with very small band overlap
or semiconductors with very small band gap are cooled to extremely 
low temperatures~\cite{JRK67,LEKMSS04}. 
The excitonic condensate typifies a macroscopic 
phase-coherent, insulating state, which separates   
the semimetal from the semiconductor (see Fig.~\ref{fig:PD}).

Surprisingly, to date, there is no free of doubt realization of 
the excitonic insulator (EI) state in nature. Nowadays experiments 
report data, however, which strongly support the theoretical predictions 
of the EI phase. Along this line, experiments on coupled quantum-well 
structures, e.g., have shown unusual properties which were 
inferred as indications
of excitonic condensation~\cite{BLIGC02}. 
Temperature dependent  angle-resolved photoelectron 
spectroscopy (ARPES) on $1T$--$\mathrm{TiSe_2}$ transition-metal 
dichalcogenides are in favor of the EI scenario as driving force 
for the observed charge-density-wave 
transition~\cite{CMCBDGBA07}.
X-ray photoemission spectroscopy and ARPES on 
quasi one-dimensional (1D) $\mathrm{Ta_2 Ni Se_5}$ reveal  
that the ground state can be viewed as EI state between 
the Ni $3d$--Se $4p$ hole and the Ta $5d$ electron~\cite{WSTMANTKNT09}. 
Further real-system candidates for the EI state are pressure-sensitive
rare-earth chalcogenides, such as mixed-valence 
$\mathrm{TmSe_{0.45}Te_{0.55}}$. For this compounds electrical and thermal
(transport) measurements indicate exciton condensation, at temperatures 
below 20~K in the pressure range between 
5 and 11 kbar~\cite{NW90}.

Also from the theoretical side the existence of the EI is still 
controversial. Most of the early mean-field approaches  
work with an effective-mass Mott-Wannier-type exciton model
and exploit the analogy to the BCS theory of 
superconductivity~\cite{Cl65} 
(for a more recent calculation of the phase diagram see 
Refs.~\onlinecite{BF06,KSH08}). Here the major problem is that 
the excitonic phases (excitonic gas, EI) turn out to be  
unstable against a metallic electron-hole 
liquid~\cite{HS84,BFR07}. At present, Falicov-Kimball-type 
models seem to be the most promising candidates for realizing 
collective exciton phases. This particularly holds for the 
extended Falicov-Kimball model 
(EFKM), which includes a direct $f$-$f$ electron hopping term, 
that---having again the Tm[Se,Te] system in mind---is certainly more 
realistic than entirely localized $f$ electrons. 
By means of unbiased constrained path Monte Carlo (CPMC) 
simulations the EFKM has been proven to exhibit critical 
excitonic correlations (an EI ground state) in 
case of 1D (2D)~\cite{Ba02b,BGBL04}. 
Subsequent Hartree-Fock calculations 
yield the ground-state phase diagram of the 2D EFKM  in 
excellent agreement with the CPMC data~\cite{Fa08}, 
supporting the applicability of such mean-field approaches 
also in the 3D situation~\cite{Fa08,SC08}. For the 3D EFKM, the 
existence of the EI phase was corroborated by more 
sophisticated slave-boson approaches~\cite{Br08,ZIBF10}. 

Assuming that $f$-$c$ electron coherence may lead to an 
EI phase in the EFKM, the properties 
of the excitonic state should be explored in more detail.  
In this regard, the anticipated `BCS-BEC crossover'
scenario~\cite{Le80,BF06,KSH08}, connecting the physics 
of BCS superconductivity with that of Bose-Einstein condensates
(BECs), is of vital importance. Calculating the Frenkel-type 
exciton propagator within a random phase approximation scheme, the existence 
of excitonic bound-states has been established for the EFKM 
also above $T_c$, on the semiconductor side of the semiconductor-semimetal 
transition~\cite{IPBBF08}. No bound-states   
were found on the semi-metallic side. Accordingly the
condensation process should differ by its nature:
While `exciton' formation and condensation simultaneously
take place on the semi-metallic (BCS) side, preformed
excitons will condense on the semiconducting (BEC) side
as the temperature is lowered (cf. Fig.~\ref{fig:PD}). 
Looking at the EI order parameter $\Delta$ only, it seems 
difficult to examine the BCS-BEC crossover. The gap equation
for $\Delta$ will of course not discriminate between both  
regimes. Photoemission spectroscopy, on the other hand,
will directly probe the elementary excitations and energy
dispersion and therefore provides extremely useful information 
about the BCS-BEC crossover. This has been shown quite recently 
in the context of ultracold (atomic) Fermi gases~\cite{SGJ08}.

In this work, we will follow this perspective and examine the 
EI phase in terms of the EFKM particularly with regard to 
a BCS-BEC crossover. To this end, we analyze the equilibrium  
and spectral properties of the model at zero temperature, 
using the so-called projective renormalization 
method (PRM)~\cite{HB06,HSB08}. This technique has 
already been successively applied to a great variety 
of many-body problems~\cite{HB05,SB09}.
Here we calculate and discuss the photoemission spectra
of the EFKM in order to probe the signatures of the excitonic 
condensate.  The paper is organized as follows.
Section~II introduces the EFKM. The theoretical approach
is outlined in Sec.~III, where the general concept 
of the PRM is resumed in Sec.~III.~A and  
explicit expressions for the renormalization differential equations,
particle number expectation values, correlation functions    
and single-particle spectral functions   
are given in Sec.~III.~B.  Section IV presents the 
corresponding numerical results. Our main conclusions
can be found in Sec.~V.

\section{Extended Falicov-Kimball model}
The Hamiltonian for the EFKM is written
\begin{equation}\label{Hami}
\mathcal{H}=\sum_{\mathbf{k}}\bar{\varepsilon}^c_{\mathbf{k}}c^\dagger_{\mathbf{k}}c^{}_{\mathbf{k}}
+\sum_{\mathbf{k}}\bar{\varepsilon}^f_{\mathbf{k}}f^\dagger_{\mathbf{k}}f^{}_{\mathbf{k}}
+\sum_{i}Un^c_in^f_i\,,
\end{equation}
where   
$c^\dagger_{\mathbf{k}}$ ($c^{}_{\mathbf{k}}$) and 
$f^\dagger_{\mathbf{k}}$ ($f^{}_{\mathbf{k}}$)
are the creation (annihilation) operators in momentum ($\mathbf{k}$-) 
space of spinless  $c$- and $f$-electrons, respectively,
and $n^c_i$ and $n^f_i$ are the corresponding occupation numbers
in real space. The Fourier-transformed fermionic operators are defined
via $\eta^\dagger_{\mathbf{k}}=\frac{1}{\sqrt{N}}\sum_i 
\eta^\dagger_{i}e^{i{\mathbf k}{\mathbf R}_i}$, where $\eta=c,f$, and  
the $\eta$-fermion dispersion is
 \begin{equation}\label{disp}
\bar{\varepsilon}^{\eta}_{\mathbf{k}}=\varepsilon_{}^{\eta}-t^{\eta}_{}
\gamma^{}_{\mathbf k}-\mu
\end{equation}
with on-site energy $\varepsilon^{\eta}_{}$. In Eq.~\eqref{disp}, $\mu$ 
denotes the chemical potential. In the tight-binding limit, on a 
D-dimensional hypercubic lattice, we have 
$\gamma^{}_{\mathbf k}=2 \sum_{d=1}^D \cos k_d$. 
The sign of $t^ct^f$  determines whether we deal with 
a direct ($t^ct^f <0$) or indirect ($t^ct^f>0$) band-gap situation. 
Usually, the $c$-electrons are considered to be `light' and
their hopping integral is taken to be the unit of energy ($t^c=1$), 
while the $f$-electrons are `heavy', i.e., $|t^f|< 1$. 
For $t^f\equiv 0$ (dispersionless $f$ band), 
the local $f$-electron number is strictly conserved~\cite{SC08}.     
The third term in the Hamiltonian~\eqref{Hami} represents the 
Coulomb interaction between $c$ and $f$ electrons at the same 
lattice site. Hence, if the $c$ and $f$ bands are degenerate, 
$\varepsilon^{c}=\varepsilon^f$ and $t^c=t^f$, 
the EFKM reduces to the standard Hubbard model. 

In order to address the formation of the EI state in the EFKM, 
we look for a non-vanishing excitonic expectation value 
$\langle c^\dagger f^{}\rangle$, indicating a kind of 
spontaneous symmetry breaking due the pairing of $c$ 
electrons ($t^c>0$) with $f$ holes ($t^f <0$). 
This is quite similar to the problem of electronic ferroelectricity, 
where $\langle c^\dagger f^{}\rangle\neq 0$ causes electrical 
polarizability without an interband transition driving field, 
provided the $c$ and $f$ states have different parity~\cite{Ba02b,SC08}.    
Thereby, depending on the sign of $t_f$ (direct or indirect gap),
ferro- or antiferroelectric phases may exist. 

To proceed, we introduce two-particle interaction operators 
in momentum space,
\begin{equation}\label{aop}
a^{}_{\mathbf{k}_1\mathbf{k}_2\mathbf{k}_3}=c^\dagger_{\mathbf{k}_1}
c^{}_{\mathbf{k}_2}
f^\dagger_{\mathbf{k}_3}f^{}_{\mathbf{k}_1+\mathbf{k}_3-\mathbf{k}_2}\,,
\end{equation}
and rewrite the EFKM Hamiltonian~\eqref{Hami} in a normal-ordered 
form~\cite{Ke06}: 
\begin{eqnarray}
\label{HFourierNO}
\mathcal{H}&=&\sum_{\mathbf{k}}\varepsilon^c_{\mathbf{k}}:c^\dagger_{\mathbf{k}}c^{}_{\mathbf{k}}:
+\sum_{\mathbf{k}}\varepsilon^f_{\mathbf{k}}:f^\dagger_{\mathbf{k}}f^{}_{\mathbf{k}}:\\
&-&\sum_{\mathbf{k}}\left(\mit{\Delta}:f^\dagger_{\mathbf{k}}c^{}_{\mathbf{k}}:+\textrm{H.c.}\right)
+\frac{U}{N}\sum_{\mathbf{k}_1\mathbf{k}_2\mathbf{k}_3}
:a^{}_{\mathbf{k}_1\mathbf{k}_2\mathbf{k}_3}:\,,\nonumber
\end{eqnarray}
where 
\begin{equation}
 \mit{\Delta}  = \frac{U}{N}\sum_{\mathbf{k}}{d}^{}_{\mathbf{k}}
\label{Delta}
\end{equation}
with ${d}^{}_{\mathbf{k}}=\langle c^\dagger_{\mathbf{k}}f^{}_{\mathbf{k}}\rangle$, 
plays the role of the EI order parameter. Note that choosing 
the normal-ordered representation of operators, the symmetry 
of the Hamiltonian is explicitly broken, and iterating  
the self-consistency equation derived below will readily 
give  (meta-) stable solutions~\cite{KW06}. 
In the Hamiltonian~\eqref{HFourierNO}, the on-site energies were  
shifted by a Hartree term,
\begin{equation}
  \varepsilon^{c(f)}_{\mathbf{k}}= \bar{\varepsilon}^{c(f)}_{\mathbf{k}}
+U\langle n^{f(c)}\rangle\,,
\label{hartreeshift}
\end{equation}
where $\langle n^\eta\rangle=\frac{1}{N}\sum_{\mathbf{k}} 
\langle \eta^\dagger_{\mathbf{k}}\eta^{}_{\mathbf{k}}\rangle$ 
are the particle number densities of $c$ or $f$ electrons for  
a system  with $N$ lattice sites. In what follows, 
we consider the half-filled band case, i.e., we fix the 
total electron density $n=\langle n^c\rangle+\langle n^f\rangle=1$. 
\section{Theoretical Approach}
\subsection{Projector-based renormalization method}
The PRM was recently developed with the aim to diagonalize 
many-particle systems~\cite{BHS02}. 
One of the main advantages of the method is to find broken 
symmetry solutions of phase transitions~\cite{SHB09a}. 
For example,  in Ref.~\onlinecite{SB09}
 the method was successfully applied to the $t-J$ model
 in order to study superconducting $d$-wave solutions for cuprates. 

The PRM starts from the decomposition of a given many-particle Hamiltonian into an `unperturbed' part  $\mathcal{H}_0$ and into a `perturbation' $\mathcal{H}_1$, where the unperturbed part $\mathcal{H}_0$ should be solvable. 
Suppose all diagonal matrix elements of $\mathcal{H}_1$ between eigenvectors of  $\mathcal{H}_0$ vanish, 
the part $\mathcal{H}_1$ accounts for all transitions between the eigenstates of $\mathcal{H}_0$ with nonzero transition energies. Then, the first goal of the PRM is to transform the initial Hamiltonian into an effective Hamiltonian $\mathcal{H}_\lambda$ which contains no longer transition operators with energies larger than 
some chosen cutoff $\lambda$. Thereby, the Hamiltonian $\mathcal{H}_\lambda$ is formally obtained by applying a unitary transformation
\begin{equation}\label{Hlda1}
\mathcal{H}_\lambda=e^{\mathcal{X}_\lambda}\mathcal{H}e^{-\mathcal{X}_\lambda}.
\end{equation}
The transformed Hamiltonian $\mathcal{H}_\lambda$,
which has the same eigenspectrum as the original Hamiltonian $\mathcal H$,
 can again be decomposed into two parts
\begin{equation}\label{Hlda2}
\mathcal{H}_\lambda=\mathcal{H}_{0,\lambda}+\mathcal{H}_{1,\lambda}\,.
\end{equation}
Due to construction, all matrix elements $\langle n_\lambda|\mathcal{H}_{1,\lambda}|m_\lambda \rangle$ of  ${\cal H}_{1,\lambda}$ with energy differences $|E^\lambda_n-E^\lambda_m|>\lambda$ should vanish, i.e.~$\langle n_\lambda| \mathcal{H}_{1,\lambda}|m_\lambda \rangle=0$, where $E^\lambda_n$ and $|n_\lambda\rangle$ 
are the new renormalized eigenvalues and eigenstates of ${\cal H}_{0,\lambda}$. Note that neither $|n_\lambda\rangle$ nor $|m_\lambda \rangle$ have to be low-energy eigenstates of ${\cal H}_{0,\lambda}$. To ensure hermiticity of ${\cal H}_{\lambda}$, the generator $\mathcal{X}_{\lambda}$ of the unitary transformation has to satisfy $\mathcal{X}^\dagger_{\lambda}=-\mathcal{X}_{\lambda}$.

A crucial idea for the elimination procedure in the PRM is to introduce generalized projection operators $ {\mathbf P}_{\lambda}$ and $ {\mathbf Q}_{\lambda}= {\mathbf 1}- {\mathbf P}_{\lambda}$. Here $ {\mathbf P}_{\lambda}$ is defined by 
\begin{eqnarray}
  \label{G5}
  {\mathbf P}_{\lambda}\mathcal{A}=\sum_{\substack{{m,n}\\{|E_n^{\lambda}-E_m^{\lambda}|\le \lambda}}}
  |n_\lambda \rangle \langle m_\lambda| \, \langle n_\lambda| \mathcal{A}  |m_\lambda \rangle ,
\end{eqnarray}
applied on any operator variable ${\mathcal A}$ of the Hilbert space of the system. Note that in expression~(\ref{G5}) only states $|n_\lambda\rangle$
and $|m_\lambda \rangle$ satisfying $\left| E_n^{\lambda} - E_m^{\lambda} \right| \leq \lambda$ contribute
to the transition matrix. Thus, $ {\mathbf P}_{\lambda}$ projects on the low energy 
transitions of ${\mathcal A}$, whereas the  orthogonal complement ${\mathbf Q}_{\lambda}$ projects on the high-energy transitions of $\mathcal{A}$. To find an appropriate generator $\mathcal{X}_{\lambda}$
for the unitary transformation from $\mathcal H$ to ${\mathcal H}_\lambda$, the obvious relation 
\begin{eqnarray}
  \label{G81}
  {\mathbf Q}_{\lambda}{\cal H}_{\lambda} = 0
\end{eqnarray}
has to be fulfilled.

In the original version of the PRM~\cite{HB06},  the elimination procedure is performed step-wise.  
Suppose $\Lambda$ is the largest transition energy of the original Hamiltonian $\mathcal H$,  in the first elimination step 
all transitions in an energy shell of width $\Delta \lambda$ between $\Lambda$
and $\Lambda-\Delta\lambda$ will be removed. The subsequent steps remove, roughly speaking, all transitions 
in the next shell of width $\Delta \lambda$ between $\Lambda-\Delta\lambda$ and $\Lambda-2\Delta\lambda$, and so on. The unitary transformation 
for the intermediate step from a
cutoff $\lambda$ to the new cutoff $\lambda-\Delta\lambda$ reads
\begin{eqnarray}
\label{G9}
{\mathcal H}_{\lambda-\Delta\lambda} = e^{\mathcal{X}_{\lambda,\Delta\lambda}}\,{\mathcal H}_\lambda \, e^{-\mathcal{X}_{\lambda,\Delta\lambda}}\,.
\end{eqnarray}
Here, the generator $\mathcal{X}_{\lambda,\Delta\lambda}$ has to fulfill the requirement
\begin{eqnarray}
  \label{G10}
  {\mathbf Q}_{\lambda-\Delta\lambda}{\mathcal H}_{\lambda-\Delta\lambda} = 0,
\end{eqnarray}
in analogy to~(\ref{G81}). Thus  ${\mathcal H}_{\lambda-\Delta\lambda}$ has no matrix elements 
that connect eigenstates of ${\mathcal H}_{0,\lambda-\Delta\lambda}$ with energy differences larger 
than $\lambda-\Delta\lambda$. 

By help of Eqs.~\eqref{G9} and~\eqref{G10}, the generator $\mathcal{X}_{\lambda,\Delta\lambda}$ 
can easily be constructed in perturbation theory with respect to ${\mathcal H}_{1, \lambda}$.  
Up to first order it reads
\begin{eqnarray}
\label{13}
\label{X}
\mathcal{X}_{\lambda,\Delta\lambda} = \frac{1}{{\mathbf L}_{0, \lambda }} 
{\mathbf Q}_{\lambda - \Delta \lambda} {\mathcal H}_{1, \lambda}
=
 {\mathbf Q}_{\lambda - \Delta \lambda}\mathcal{X}_{\lambda,\Delta\lambda}\,. 
\end{eqnarray}
Here, ${\mathbf L}_{0, \lambda }$ is the Liouville superoperator of the unperturbed Hamiltonian 
${\mathcal H}_{0,\lambda}$, which is defined by the commutator of ${\mathcal H}_{0,\lambda}$
with any operator ${\mathcal A}$ on which ${\mathbf L}_{0, \lambda }$ is applied, i.e.~${\mathbf L}_{0, \lambda }{\mathcal A} =
 [{\mathcal H}_{0,\lambda} , {\mathcal A}]$. Note, however,  that 
the generator $\mathcal{X}_{\lambda,\Delta\lambda}$ is not completely fixed by Eqs.~\eqref{G9} and~\eqref{G10}. In fact, only the part ${\mathbf Q}_{\lambda-\Delta\lambda}\mathcal{X}_{\lambda,\Delta\lambda}$ is determined by equation \eqref{G10}. 
The part  ${\mathbf P}_{\lambda-\Delta\lambda}\mathcal{X}_{\lambda,\Delta\lambda}$
with only low-energy transitions can still be chosen arbitrarily and was set identical to zero in 
Eq.~\eqref{X}. Any physical quantities, which is evaluated in the framework of the PRM, 
is independent of a particular  choice of  ${\mathbf P}_{\lambda-\Delta\lambda}\mathcal{X}_{\lambda,\Delta\lambda}$~\cite{HSB08}. 
This freedom can be used to derive a continuous version of the method. Thereby, the low excitation part 
${\mathbf P}_{\lambda-\Delta\lambda}\mathcal{X}_{\lambda,\Delta\lambda}$ is chosen proportional to
$\Delta \lambda$,  which allows to derive differential equations for the $\lambda$-dependence of the parameter 
values in the Hamiltonian during the renormalization procedure. 
As in the discrete version~\cite{HB06}, also in the continuous version the elimination 
starts from the original model and proceeds until $\lambda=0$. At this point, all 
transitions operators from ${\mathcal H}_1$ have been used up and the final Hamiltonian 
is diagonal or at least quasi-diagonal which allows to evaluate expectation values. Note that the 
parameters  of the renormalized Hamiltonian  depend on the parameter values of the original model $\mathcal H$.

To evaluate expectation values of operators $\mathcal A$, 
formed with the full Hamiltonian,
we have to apply the unitary transformation to $\mathcal A$ as well,
\begin{eqnarray}
\label{erw}
 \langle {\mathcal A} \rangle =
\frac{{\mbox{Tr}} {\mathcal A} e^{-\beta {\mathcal H}}}{ {\mbox{Tr}} e^{-\beta {\mathcal H}}} = 
\langle {\mathcal A}(\lambda) \rangle_{{\mathcal H}_\lambda}=
 \langle \tilde{{\mathcal A}}\rangle_{\tilde{{\mathcal H}}} \, , 
\end{eqnarray}
where we define 
${\mathcal A}(\lambda) = e^{\mathcal{X}_\lambda}{\mathcal A}e^{-\mathcal{X}_\lambda}$, $\tilde{{\mathcal A}}= 
{\mathcal A}(\lambda \rightarrow 0)$, and $\tilde{\mathcal H}= {\mathcal H}_{\lambda \rightarrow 0}$. 
Thus additional renormalization equations 
are required for  ${\mathcal A}(\lambda)$. 
\subsection{Application to the EFKM}

\subsubsection{Renormalization equations}

In order to derive the renormalization equations for the parameters of the Hamiltonian, we first decompose the original 
Hamiltonian $\mathcal H$ into two parts
\begin{equation}\label{Horiginal}
\mathcal{H}=\mathcal{H}_0+\mathcal{H}_1\,,
\end{equation}
where
\begin{eqnarray}\label{H0}
\mathcal{H}_0&=&\sum_{\mathbf{k}}\varepsilon^c_{\mathbf{k}}:c^\dagger_{\mathbf{k}}c^{}_{\mathbf{k}}:
+\sum_{\mathbf{k}}\varepsilon^f_{\mathbf{k}}:f^\dagger_{\mathbf{k}}f^{}_{\mathbf{k}}:\nonumber\\
&+&\sum_{\mathbf{k}}\left(\mit{\Delta}:f^\dagger_{\mathbf{k}}c^{}_{\mathbf{k}}:+\textrm{H.c.}\right)\,,
\end{eqnarray}
and
\begin{equation}\label{H1}
\mathcal{H}_1=\frac{U}{N}\sum_{\mathbf{k}_1\mathbf{k}_2\mathbf{k}_3}
:a^{}_{\mathbf{k}_1\mathbf{k}_2\mathbf{k}_3}:\,.
\end{equation}
Note again that the perturbation ${\mathcal H}_1$ only contains the fluctuating operator part of the Coulomb repulsion $\propto U$. 
Following the ideas of the PRM approach, we make the following {\em ansatz} for the renormalized Hamiltonian 
${\mathcal H}_\lambda$ after all transitions with energies larger than $\lambda$
are integrated out: 
\begin{equation}\label{Hlda}
\mathcal{H_\lambda}=\mathcal{H}_{0,\lambda}+\mathcal{H}_{1,\lambda}
\end{equation}
with
\begin{eqnarray}
\label{H0l}
\mathcal{H}_{0,\lambda}&=&\sum_{\mathbf{k}}\varepsilon^c_{\mathbf{k},\lambda}:c^\dagger_{\mathbf{k}}c^{}_{\mathbf{k}}:   
+\sum_{\mathbf{k}}\varepsilon^f_{\mathbf{k},\lambda}:f^\dagger_{\mathbf{k}}f^{}_{\mathbf{k}}:+E_\lambda
\nonumber\\
&&+\sum_{\mathbf{k}}\left(\Delta^{}_{\mathbf{k},\lambda}:f^\dagger_{\mathbf{k}}c^{}_{\mathbf{k}}:+\textrm{H.c.}\right)\,,\\
\label{H1l}
\mathcal{H}_{1,\lambda}&=&\frac{1}{N}\mathbf{P}_\lambda \sum_{\mathbf{k}_1\mathbf{k}_2\mathbf{k}_3}
U^{}_{\mathbf{k}_1\mathbf{k}_2\mathbf{k}_3,\lambda}\,:a^{}_{\mathbf{k}_1\mathbf{k}_2\mathbf{k}_3}:.
\end{eqnarray}
Here, $\mathbf{P}_\lambda$ projects on all low-energy transitions with respect to the 
unperturbed Hamiltonian $\mathcal{H}_{0,\lambda}$ which are smaller than $\lambda$. 
Due to renormalization 
all prefactors in Eqs. (\ref{H0l}), \eqref{H1l}  may now depend on the wave vector ${\mathbf k}$ and on the 
energy cutoff $\lambda$.  
The quantity $E_\lambda$ is an energy shift which enters during the renormalization
procedure. 
In order to evaluate the action of the superoperator $\mathbf{P}_\lambda$ on the interaction operator in $\mathcal{H}_{1,\lambda}$ 
one has to decompose the fluctuation operators $:a^{}_{\mathbf{k}_1\mathbf{k}_2\mathbf{k}_3}:$
into eigenmodes of ${\mathcal H}_{0,\lambda}$. Obviously, the diagonalization of
  ${\mathcal H}_{0,\lambda}$ requires an additional unitary transformation. However, for the values of $U$, used 
  in the numerical evaluation below, the mixing parameter $\Delta_{{\mathbf k},\lambda}$ in Eq.~\eqref{H0l} turns out to be always small compared to the energy difference  $|\varepsilon_{{\mathbf k}, \lambda}^c - \varepsilon_{{\mathbf k}, \lambda}^f|$. This follows from the Hartree shifts of the one-particle energies in Eq.~\eqref{hartreeshift}. Thus, using as approximation
 ${\mathbf L}_{0,\lambda} c_{\mathbf  k}^\dagger = \varepsilon_{\mathbf k}^c  c_{\mathbf  k}^\dagger$  
 and 
 ${\mathbf L}_{0,\lambda} f_{\mathbf  k}^\dagger = \varepsilon_{\mathbf k}^f  f_{\mathbf  k}^\dagger$,
 we can conclude 
 \begin{equation}
\label{H1lda}
\mathcal{H}_{1,\lambda}=\frac{1}{N}\sum_{\mathbf{k}_1\mathbf{k}_2\mathbf{k}_3}\Theta(\lambda-|\tilde{\eta}^{}_{{\mathbf k}_1{\mathbf k}_2{\mathbf k}_3,\lambda}|)\, U^{}_{\mathbf{k}_1\mathbf{k}_2\mathbf{k}_3,\lambda}
:a^{}_{\mathbf{k}_1\mathbf{k}_2\mathbf{k}_3}:\,,
\end{equation}
where 
\begin{equation}
\tilde{\eta}^{}_{{\mathbf k}_1{\mathbf k}_2{\mathbf k}_3,\lambda}=
\varepsilon^c_{{\mathbf k}_1,\lambda}-\varepsilon^c_{{\mathbf k}_2,\lambda}
+\varepsilon^f_{{\mathbf k}_3,\lambda}-\varepsilon^f_{{\mathbf k}_1+{\mathbf k}_3-{\mathbf k}_2,\lambda}
\end{equation}
is the approximate excitation energy of $:a^{}_{\mathbf{k}_1\mathbf{k}_2\mathbf{k}_3}:$, i.e.
\begin{eqnarray}
\label{L0}
{\mathbf L}_{0, \lambda}\, :a^{}_{\mathbf{k}_1\mathbf{k}_2\mathbf{k}_3}:  = 
\tilde{\eta}^{}_{{\mathbf k}_1{\mathbf k}_2{\mathbf k}_3,\lambda}  :a^{}_{\mathbf{k}_1\mathbf{k}_2\mathbf{k}_3}:\,.
\end{eqnarray}
The $\Theta$-function in Eq.~\eqref{H1lda} ensures that  
only transitions with excitation energies smaller than $\lambda$
remain in ${\mathcal H}_{1,\lambda}$.

By integrating out all transitions between the cutoff $\Lambda$ of the original model and $\lambda=0$,
all parameters of the original model will become renormalized. To find their $\lambda$-dependence,
we derive renormalization equations for the parameters 
$\varepsilon_{{\mathbf k},\lambda}^c, 
\varepsilon_{{\mathbf k},\lambda}^f, \Delta_{{\mathbf k}, 
\lambda}$, and $U^{}_{\mathbf{k}_1\mathbf{k}_2\mathbf{k}_3,\lambda}$. 
The initial parameter values are determined by the original model ($\lambda= \Lambda$):
\begin{eqnarray}
\varepsilon^c_{\mathbf{k},\Lambda}&=&\varepsilon^c_{\mathbf{k}}\, , \quad
\Delta^{}_{\mathbf{k},\Lambda}=\mit{\Delta}\,,\\
\varepsilon^f_{\mathbf{k},\Lambda}&=&\varepsilon^f_{\mathbf{k}}\, , \quad
U^{}_{\mathbf{k}_1\mathbf{k}_2\mathbf{k}_3,\Lambda}=U\,.
\label{IniH}
\end{eqnarray}
Note that the energy shift $E_\lambda$ in ${{\mathcal H}_{0, \lambda}}$ has no effect on expectation values 
and will be left out in what follows.

Next we have to construct the generator $\mathcal{X}_{\lambda, \Delta \lambda}$ of 
transformation \eqref{G9}. Using relation \eqref{L0}, the high transition energy part reads 
in lowest order perturbation theory according to Eqs.~\eqref{13}
\begin{eqnarray}
\label{24}
{\mathbf Q}_{\lambda - \Delta \lambda}\mathcal{X}_{\lambda, \Delta \lambda} &=& 
\frac{1}{N}\sum_{{\mathbf k}_1{\mathbf k}_2{\mathbf k}_3}
\frac{U^{}_{\mathbf{k}_1\mathbf{k}_2\mathbf{k}_3,\lambda}}
 {\tilde{\eta}^{}_{{\mathbf k}_1{\mathbf k}_2{\mathbf k}_3,\lambda}} \, 
\big(1- \Theta_{{\mathbf k}_1{\mathbf k}_2{\mathbf k}_3, \lambda- \Delta \lambda }
\big)\nonumber\\ 
&&\hspace*{1.5cm}\times\Theta_{{\mathbf k}_1{\mathbf k}_2{\mathbf k}_3, \lambda}
:a^{}_{\mathbf{k}_1\mathbf{k}_2\mathbf{k}_3}:\,,
\end{eqnarray}
where we have defined
$
\Theta_{{\mathbf k}_1{\mathbf k}_2{\mathbf k}_3, \lambda} =
\Theta(\lambda-|\tilde{\eta}^{}_{{\mathbf k}_1{\mathbf k}_2{\mathbf k}_3,\lambda}|)
$. Here the product of the two $\Theta$-function assures that only excitations between 
$\lambda - \Delta \lambda$ and $\lambda$ are eliminated by the unitary transformation \eqref{G9}.
As mentioned before, in the present approach we prefer 
to use a continuous version of the PRM approach which is based on the choice of the orthogonal complement 
part  ${\mathbf P}_{\lambda - \Delta \lambda}\mathcal{X}_{\lambda, \Delta \lambda}$
of the generator. Thereby, ${\mathbf P}_{\lambda - \Delta \lambda}\mathcal{X}_{\lambda, \Delta \lambda}$
is chosen proportional to $\Delta \lambda$,  which means that 
${\mathbf Q}_{\lambda - \Delta \lambda}\mathcal{X}_{\lambda, \Delta \lambda}$ can be neglected 
in  the limit $\Delta \lambda \rightarrow 0$~\cite{HSB08}. With 
$\mathcal{X}_{\lambda, \Delta \lambda} \approx{\mathbf P}_{\lambda - \Delta \lambda}\mathcal{X}_{\lambda, \Delta \lambda}$,  
the following operator form for the generator can be used  
\begin{eqnarray}
\label{XDlda}
\mathcal{X}_{\lambda,\Delta\lambda}&=& 
\frac{\Delta \lambda}{N}\sum_{\mathbf{k}_1\mathbf{k}_2\mathbf{k}_3}
\tilde{\alpha}^{}_{\mathbf{k}_1\mathbf{k}_2\mathbf{k}_3,\lambda} \,
\Theta_{{\mathbf k}_1{\mathbf k}_2{\mathbf k}_3, \lambda- \Delta \lambda}\nonumber\\ 
&&\hspace*{1.5cm} 
\times\Theta_{{\mathbf k}_1{\mathbf k}_2{\mathbf k}_3, \lambda} \,
:a^{}_{\mathbf{k}_1\mathbf{k}_2\mathbf{k}_3}:\,,
\end{eqnarray}
where the operators are taken over from expression \eqref{24}.
Note that the two $\Theta$-functions guarantee that expression \eqref{XDlda} corresponds to 
the generator part with low energy excitations only.  For the  coefficients 
$\tilde{\alpha}^{}_{\mathbf{k}_1\mathbf{k}_2\mathbf{k}_3,\lambda}$
we make the following {\em ansatz}:
\begin{eqnarray}
\label{alpha}
{\tilde\alpha}^{}_{\mathbf{k}_1\mathbf{k}_2\mathbf{k}_3, \lambda} 
&=&\frac{\tilde{\eta}^{}_{{\mathbf k}_1{\mathbf k}_2{\mathbf k}_3,\lambda}}
{\kappa(\lambda-|\tilde{\eta}_{{\mathbf k}_1{\mathbf k}_2{\mathbf k}_3,\lambda}|)^2}\, U^{}_{\mathbf{k}_1\mathbf{k}_2\mathbf{k}_3,\lambda}
\end{eqnarray}
which is an appropriate choice in the continuous version of the 
PRM~\cite{HSB08}. 
The constant $\kappa$ in (\ref{alpha}) denotes an energy constant to ensure that the 
parameter $\tilde{\alpha}^{}_{\mathbf{k}_1\mathbf{k}_2\mathbf{k}_3, \lambda}$ 
has the correct dimension of an inverse energy. 
At first glance, one might expect that 
$\tilde{\alpha}_{\mathbf{k}_1\mathbf{k}_2\mathbf{k}_3, \lambda}$ diverges at 
$\lambda=|\tilde{\eta}_{{\mathbf k}_1{\mathbf k}_2{\mathbf k}_3,\lambda}|$. Instead it vanishes exponentially at this point which follows from 
the renormalization equation for 
$U^{}_{\mathbf{k}_1\mathbf{k}_2\mathbf{k}_3,\lambda}$, given below.

Our aim is to derive renormalization equations for renormalized Hamiltonian. The transformation 
\eqref{G9} relates the Hamiltonian ${\mathcal H}_\lambda$ at cutoff $\lambda$ to that at the reduced cutoff $\lambda - \Delta \lambda$. With Eq.~\eqref{XDlda} 
 one finds in the limit $\Delta \lambda \rightarrow 0$:
\begin{equation}
\frac{d{\mathcal H}_\lambda}{d\lambda} =-
\frac{1}{N}\sum_{\mathbf{k}_1\mathbf{k}_2\mathbf{k}_3}
\tilde{\alpha}^{}_{\mathbf{k}_1\mathbf{k}_2\mathbf{k}_3,\lambda} 
\Theta_{{\mathbf k}_1{\mathbf k}_2{\mathbf k}_3, \lambda} 
[ :a^{}_{\mathbf{k}_1\mathbf{k}_2\mathbf{k}_3}:,   {\mathcal H}_{\lambda} ]\,.
\end{equation}
 Note that the evaluation of the commutator also leads to new operators which are 
 not present in the {\em ansatz}~\eqref{Hlda} for ${\mathcal H}_\lambda$. Therefore an additional factorization has to be used.  Comparison with the generic derivation of \eqref{Hlda}
 leads to the following set of coupled renormalization equations which describe the 
 $\lambda$-dependent renormalization of the parameters of 
${\mathcal H}_\lambda$:
 
 \begin{align}\label{ReEqs1}
\frac{d\varepsilon^c_{\mathbf{k},\lambda}}{d\lambda}=
-&\frac{1}{N^2}\sum_{\mathbf{k}_1\mathbf{k}_2}U^{}_{\mathbf{k}_1\mathbf{k}\mathbf{k}_2,\lambda}
\tilde{\alpha}^{}_{\mathbf{k}\mathbf{k}_1,\mathbf{k}_1+\mathbf{k}_2-\mathbf{k},\lambda}
(1-\langle n^c_{\mathbf{k}_1}\rangle)\nonumber\\
&\times(\langle n^f_{\mathbf{k}_1+\mathbf{k}_2-\mathbf{k}}\rangle-\langle n^f_{\mathbf{k}_2}\rangle)\nonumber\\
-&\frac{1}{N^2}\sum_{\mathbf{k}_1\mathbf{k}_2}U^{}_{\mathbf{k}\mathbf{k}_1\mathbf{k}_2,\lambda}
\tilde{\alpha}^{}_{\mathbf{k}_1\mathbf{k},\mathbf{k}+\mathbf{k}_2-\mathbf{k}_1,\lambda}
\langle n^c_{\mathbf{k}_1}\rangle\nonumber\\
&\times(\langle n^f_{\mathbf{k}+\mathbf{k}_2-\mathbf{k}_1}\rangle-\langle n^f_{\mathbf{k}_2}\rangle)\,,
\end{align}
\begin{align}\label{ReEqs2}
\frac{d\varepsilon^f_{\mathbf{k},\lambda}}{d\lambda}=
-&\frac{1}{N^2}\sum_{\mathbf{k}_1\mathbf{k}_2}U^{}_{\mathbf{k}_1\mathbf{k}_2,\mathbf{k}-\mathbf{k}_1+\mathbf{k}_2,\lambda}
\tilde{\alpha}^{}_{\mathbf{k}_2\mathbf{k}_1\mathbf{k},\lambda}
\langle n^f_{\mathbf{k}-\mathbf{k}_1+\mathbf{k}_2}\rangle \nonumber\\&\times(\langle n^c_{\mathbf{k}_1}\rangle-\langle n^c_{\mathbf{k}_2}\rangle)\nonumber\\
-&\frac{1}{N^2}\sum_{\mathbf{k}_1\mathbf{k}_2}U^{}_{\mathbf{k}_1\mathbf{k}_2\mathbf{k},\lambda}
\tilde{\alpha}^{}_{\mathbf{k}_2\mathbf{k}_1,\mathbf{k}+\mathbf{k}_1-\mathbf{k}_2,\lambda}\nonumber\\
&\times(1-\langle n^f_{\mathbf{k}+\mathbf{k}_1-\mathbf{k}_2}\rangle)(\langle n^c_{\mathbf{k}_1}\rangle-\langle n^c_{\mathbf{k}_2}\rangle)\,,
\end{align}
\begin{align}\label{ReEqs3}
\frac{d\Delta^{}_{\mathbf{k},\lambda}}{d\lambda}=
-&\frac{1}{N}\sum_{\mathbf{k}_1}
\tilde{\alpha}_{\mathbf{k}_1\mathbf{k}\mathbf{k},\lambda}\Delta^{}_{\mathbf{k}_1,\lambda}(\langle n^{f}_{\mathbf{k}_1}\rangle
-\langle n^{c}_{\mathbf{k}_1}\rangle)\nonumber\\
-&\frac{1}{N^2}\sum_{\mathbf{k}_1\mathbf{k}_2}\Big\{U^{}_{\mathbf{k}_1\mathbf{k}_2\mathbf{k}_2,\lambda}
\tilde{\alpha}^{}_{\mathbf{k}\mathbf{k}_1\mathbf{k}_1,\lambda}
{\mit{d}}^{}_{\mathbf{k}}(1-\langle n^c_{\mathbf{k}_1}\rangle)\nonumber\\
&+U^{}_{\mathbf{k}_1\mathbf{k}\mathbf{k},\lambda}
\tilde{\alpha}^{}_{\mathbf{k}\mathbf{k}_2\mathbf{k}_2,\lambda}
{\mit{d}}^{}_{\mathbf{k}_1}\langle n^c_{\mathbf{k}}\rangle\nonumber\\
&-U^{}_{\mathbf{k}_1\mathbf{k}_2\mathbf{k},\lambda}\left[
\tilde{\alpha}^{}_{\mathbf{k}_1+\mathbf{k}-\mathbf{k}_2,\mathbf{k}_1\mathbf{k}_2,\lambda}
{\mit{d}}^{}_{\mathbf{k}_1+\mathbf{k}-\mathbf{k}_2}\langle n^f_{\mathbf{k}}\rangle
\right.\nonumber\\&\,\,\,+\left.\tilde{\alpha}^{}_{\mathbf{k}_2\mathbf{k},\mathbf{k}_1+\mathbf{k}-\mathbf{k}_2,\lambda}
{\mit{d}}^{}_{\mathbf{k}_1}(1-\langle n^f_{\mathbf{k}_1+\mathbf{k}-\mathbf{k}_2}\rangle)\right]\Big\}.
\end{align}
Here, we have defined expectation values
\begin{eqnarray}
\label{expec}
\langle n_{\mathbf k}^c\rangle \!&=&\!  \langle c_{\mathbf k}^\dagger  c_{\mathbf k}\rangle, \;\;
\langle n_{\mathbf k}^f\rangle =  \langle f_{\mathbf k}^\dagger  f_{\mathbf k}\rangle, \;\;
{d}_{\mathbf k} =  \langle c_{\mathbf k}^\dagger  f_{\mathbf k}\rangle ,
\end{eqnarray}
which are formed with the full Hamiltonian.
There is also an additional renormalization equation for the 
$\lambda$-dependent coupling $U^{}_{\mathbf{k}_1\mathbf{k}_2\mathbf{k},\lambda}$. It reads
\begin{eqnarray}
\frac{d U^{}_{\mathbf{k}_1\mathbf{k}_2\mathbf{k},\lambda}}{d\lambda} =\tilde{\eta}^{}_{\mathbf{k}_1\mathbf{k}_2\mathbf{k},\lambda}\tilde{\alpha}^{}_{\mathbf{k}_1\mathbf{k}_2\mathbf{k},\lambda}\,.
\end{eqnarray}
Integrating the whole set of differential equations with the initial values given by 
Eq.~\eqref{IniH}, the completely renormalized Hamiltonian $\tilde{\mathcal H}:=
{\mathcal H}_{\lambda \rightarrow 0}={\mathcal H}_{0, \lambda \rightarrow 0}$
is obtained 
\begin{eqnarray}
\label{ReH}
\tilde{\mathcal{H}}&=&
 \sum_{\mathbf{k}}\tilde{\varepsilon}^c_{\mathbf{k}}:c^\dagger_{\mathbf{k}}c^{}_{\mathbf{k}}:
+\sum_{\mathbf{k}}\tilde{\varepsilon}^f_{\mathbf{k}}:f^\dagger_{\mathbf{k}}f^{}_{\mathbf{k}}:\nonumber\\
&&+\sum_{\mathbf{k}}(\tilde{\Delta}^{}_{\mathbf{k}}:f^\dagger_{\mathbf{k}}c^{}_{\mathbf{k}}:+\textrm{H.c.}),
\end{eqnarray}
where the quantities with tilde sign denote the parameter values at $\lambda \rightarrow 0$. 
The final Hamiltonian \eqref{ReH} can be diagonalized by use of a Bogoliubov transformation~\cite{PVN09}:
\begin{eqnarray}\label{HFanoRe}
{\tilde{\mathcal H}}
=\sum_{{\mathbf k}}E^c_{{\mathbf k}}:\bar{c}^\dagger_{{\mathbf k}}\bar{c}^{}_{{\mathbf k}}:
+\sum_{{\mathbf k}}E^f_{{\mathbf k}}:\bar{f}^\dagger_{{\mathbf k}}\bar{f}^{}_{{\mathbf k}}:
+\tilde{E}\,.
\end{eqnarray}
Here, $  \bar{c}^\dagger_{{\mathbf k}} $ and $  \bar{f}^\dagger_{{\mathbf k}}$ 
are the new quasiparticle operators 
\begin{eqnarray}
\label{34}
  \bar{c}^\dagger_{{\mathbf k}} &=& u^{}_{{\mathbf k}}c^\dagger_{{\mathbf k}}+v^{}_{{\mathbf k}}f^\dagger_{{\mathbf k}}\,,  \\
  \bar{f}^\dagger_{{\mathbf k}} &=& -v^{}_{{\mathbf k}}c^\dagger_{{\mathbf k}}+u^{}_{{\mathbf k}}f^\dagger_{{\mathbf k}}\,,
\end{eqnarray}
with
\begin{eqnarray}
u^2_{\mathbf k}&=&\frac{1}{2}\left(1+\textrm{sgn}(\tilde{\varepsilon}^f_{\mathbf k}
-\tilde{\varepsilon}^c_{\mathbf k})\frac{\tilde{\varepsilon}^f_{\mathbf k}
-\tilde{\varepsilon}^c_{\mathbf k}}{W_{\mathbf k}}\right)\label{xu2}\,,  \\
v^2_{\mathbf k}&=&\frac{1}{2}\left(1-\textrm{sgn}(\tilde{\varepsilon}^f_{\mathbf k}
-\tilde{\varepsilon}^c_{\mathbf k})\frac{\tilde{\varepsilon}^f_{\mathbf k}
-\tilde{\varepsilon}^c_{\mathbf k}}{W_{\mathbf k}}\right)\,.
\end{eqnarray}
The quasiparticle energies are given by
\begin{eqnarray}
\label{qpe_c}
E^c_{\mathbf k}&=&\frac{\tilde{\varepsilon}^c_{\mathbf k}+\tilde{\varepsilon}^f_{\mathbf k}}{2}
-\frac{\textrm{sgn}(\tilde{\varepsilon}^f_{\mathbf k}-\tilde{\varepsilon}^c_{\mathbf k})}
{2}W_{\mathbf k}\,,\\\label{qpe_f}
E^f_{\mathbf k}&=&\frac{\tilde{\varepsilon}^c_{\mathbf k}+\tilde{\varepsilon}^f_{\mathbf k}}{2}
+\frac{\textrm{sgn}(\tilde{\varepsilon}^f_{\mathbf k}-\tilde{\varepsilon}^c_{\mathbf k})}
{2}W_{\mathbf k}\,,
\end{eqnarray}
where
\begin{eqnarray}
W_{\mathbf k}=\sqrt{(\tilde{\varepsilon}^c_{\mathbf k}-\tilde{\varepsilon}^f_{\mathbf k})^2
+4|\tilde{\Delta}^{}_{\mathbf k}|^2}.
\end{eqnarray}

\subsubsection{Expectation values}

The expectation values \eqref{expec} in the set of renormalization equations have to be evaluated 
self-consistently.  According to relation \eqref{erw}, thereby the same unitary transformation as 
for the Hamiltonian has to be used. For instance, 
following Eq.~\eqref{erw}, the expectation value $\langle n^c_{\mathbf k}\rangle$ can be expressed by 
\begin{equation}
\label{ncc}
\langle n^c_{\mathbf k}\rangle= \langle c^\dagger_{\mathbf{k}}c^{}_{\mathbf{k}}\rangle=\langle c^\dagger_{\mathbf{k}}(\lambda\rightarrow 0)c^{}_{\mathbf{k}}(\lambda\rightarrow 0)\rangle_{\tilde{\mathcal{H}}}\,,
\end{equation}
where the average on the r.h.s.  
 is formed with the fully renormalized Hamiltonian $\tilde{\mathcal{H}}$, 
 and $c^{\dagger}_{\mathbf{k}}(\lambda)$ is given by
$c^{\dagger}_{\mathbf{k}}(\lambda) =  e^{\mathcal{X}_\lambda} c^{\dagger}_{\mathbf{k}}
e^{-\mathcal{X}_\lambda}$. For the transformed operator we use as {\em ansatz}
\begin{align}
:c^\dagger_{\mathbf{k}}(\lambda):\;=&\;x^{}_{\mathbf{k},\lambda}:c^\dagger_{\mathbf{k}}:\nonumber\\
&+\frac{1}{N}\sum_{\mathbf{k}_1\mathbf{k}_2}
y^{}_{\mathbf{k}_1\mathbf{k}\mathbf{k}_2,\lambda}:c^\dagger_{\mathbf{k}_1}
f^\dagger_{\mathbf{k}_2}f^{}_{\mathbf{k}_1+\mathbf{k}_2-\mathbf{k}}:\,.\label{clda}
\end{align}
with a coherent part  $\propto x^{}_{\mathbf{k},\lambda}$ and an incoherent part 
$\propto y^{}_{\mathbf{k}_1\mathbf{k}\mathbf{k}_2,\lambda}$. The operator structure in \eqref{clda} is again taken over from the lowest order expansion of the unitary transformation. For the $\lambda$-dependent coefficients 
$x^{}_{\mathbf{k},\lambda}$ and $y^{}_{\mathbf{k}_1\mathbf{k}\mathbf{k}_2,\lambda}$ new
 renormalization equations can be derived. They read
\begin{align}
&\frac{dx^{}_{\mathbf{k},\lambda}}{d\lambda}
=-\frac{1}{N^2}\sum_{\mathbf{k}_1\mathbf{k}_2}y^{}_{\mathbf{k}_1\mathbf{k}\mathbf{k}_2,\lambda}
\tilde{\alpha}^{}_{\mathbf{k}\mathbf{k}_1,\mathbf{k}_1+\mathbf{k}_2-\mathbf{k},\lambda}\Big[
(1-\langle n^{c}_{\mathbf{k}_1}\rangle)\nonumber\\
&\quad\quad\,\,\times(\langle n^{f}_{\mathbf{k}_1+\mathbf{k}_2-\mathbf{k}}\rangle-\langle n^{f}_{\mathbf{k}_2}\rangle)
+\langle n^{f}_{\mathbf{k}_2}\rangle(1-\langle n^{f}_{\mathbf{k}_1+\mathbf{k}_2-\mathbf{k}}\rangle)\Big],\\
&\frac{dy^{}_{\mathbf{k}_1\mathbf{k}\mathbf{k}_2,\lambda}}{d\lambda}=-x^{}_{\mathbf{k},\lambda}
\tilde{\alpha}^{}_{\mathbf{k}_1\mathbf{k}\mathbf{k}_2,\lambda}.\label{ReEqsC}
\end{align}
Integration between $\Lambda$ (where $x^{}_{\mathbf{k},\Lambda}=1$ 
and $y^{}_{\mathbf{k}_1\mathbf{k}\mathbf{k}_2,\Lambda}=0$) and $\lambda =0$ leads to 
\begin{align}
:c^\dagger_{\mathbf{k}}(\lambda\rightarrow 0)&:=\tilde{x}^{}_{\mathbf{k}}:c^\dagger_{\mathbf{k}}:\nonumber\\
+&\frac{1}{N}\sum_{\mathbf{k}_1\mathbf{k}_2}
\tilde{y}^{}_{\mathbf{k}_1\mathbf{k}\mathbf{k}_2}:c^\dagger_{\mathbf{k}_1}
f^\dagger_{\mathbf{k}_2}f^{}_{\mathbf{k}_1+\mathbf{k}_2-\mathbf{k}}:\,,\label{cRe}
\end{align}
from which $\langle n_{\mathbf k}^c \rangle$
is found
\begin{eqnarray}
\label{Eq:ncc}
\langle n^c_{\mathbf{k}} \rangle&=&
|\tilde{x}^{}_{\mathbf{k}}|^2\langle c^\dagger_{\mathbf{k}}c^{}_{\mathbf{k}}\rangle_{\tilde{\mathcal{H}}}\nonumber\\
&&+\frac{1}{N^2}\sum_{\mathbf{k}_1\mathbf{k}_2}\left|
\tilde{y}^{}_{\mathbf{k}_1\mathbf{k}\mathbf{k}_2}\right|^2
\langle c^\dagger_{\mathbf{k}_1}c^{}_{\mathbf{k}_1}\rangle_{\tilde{\mathcal{H}}}
\langle f^\dagger_{\mathbf{k}_2}f^{}_{\mathbf{k}_2}\rangle_{\tilde{\mathcal{H}}}\nonumber\\
&&\times\left(1-\langle f^\dagger_{\mathbf{k}_1+\mathbf{k}_2-\mathbf{k}}
f^{}_{\mathbf{k}_1+\mathbf{k}_2-\mathbf{k}}\rangle_{\tilde{\mathcal{H}}}\right)\,.
\end{eqnarray}

 The remaining  expectation values $\langle n^f_{\mathbf{k}} \rangle$ and $d^{}_{\mathbf{k}}$ can be evaluated  by using an equivalent {\em ansatz} for 
$:f^\dagger_{\mathbf{k}}(\lambda):$,
\begin{align}
:f^\dagger_{\mathbf{k}}&(\lambda):=x'_{\mathbf{k},\lambda}:f^\dagger_{\mathbf{k}}:\nonumber\\
&+\frac{1}{N}\sum_{\mathbf{k}_1\mathbf{k}_2}
y'_{\mathbf{k}_1\mathbf{k}_2,\mathbf{k}-\mathbf{k}_1+\mathbf{k}_2,\lambda}
:c^\dagger_{\mathbf{k}_1}c^{}_{\mathbf{k}_2}f^\dagger_{\mathbf{k}-\mathbf{k}_1+\mathbf{k}_2}:\,,\label{flda}
\end{align}
consisting again of a coherent and an incoherent part
with $\lambda$-dependent coefficients $x'_{\mathbf{k},\lambda}$ 
and $y'_{\mathbf{k}_1\mathbf{k}_2,\mathbf{k}-\mathbf{k}_1+\mathbf{k}_2,\lambda}$, respectively.  Their renormalization equations read
\begin{align}
&\frac{dx'_{\mathbf{k},\lambda}}{d\lambda}
=-\frac{1}{N^2}\sum_{\mathbf{k}_1\mathbf{k}_2}y'_{\mathbf{k}_1\mathbf{k}_2,
\mathbf{k}-\mathbf{k}_1+\mathbf{k}_2,\lambda}
\tilde{\alpha}^{}_{\mathbf{k}_2\mathbf{k}_1\mathbf{k},\lambda}\Big[
\langle n^{f}_{\mathbf{k}-\mathbf{k}_1+\mathbf{k}_2}\rangle\nonumber\\
&\quad\quad\quad\quad\times(\langle n^{c}_{\mathbf{k}_1}\rangle-\langle n^{c}_{\mathbf{k}_2}\rangle)
+\langle n^{c}_{\mathbf{k}_2}\rangle(1-\langle n^{c}_{\mathbf{k}_1}\rangle)\Big]\,,\\
&\frac{dy'_{\mathbf{k}_1\mathbf{k}_2,\mathbf{k}-\mathbf{k}_1+\mathbf{k}_2,\lambda}}{d\lambda}=-x'_{\mathbf{k},\lambda}
\tilde{\alpha}^{}_{\mathbf{k}_1\mathbf{k}_2,\mathbf{k}-\mathbf{k}_1+\mathbf{k}_2,\lambda}\,,\label{ReEqsF}
\end{align}
where the initial values are $x'_{\mathbf{k},\Lambda}=1$ and  $y'_{\mathbf{k}_1\mathbf{k}_2,\mathbf{k}-\mathbf{k}_1+\mathbf{k}_2,\Lambda}=0$. 
Similar to Eq.~(\ref{Eq:ncc}), we are led to 
\begin{eqnarray}\label{Eq:nff}
\langle n^f_{\mathbf k} \rangle&=&
|\tilde{x}'_{\mathbf{k}}|^2\langle f^\dagger_{\mathbf{k}}f^{}_{\mathbf{k}}\rangle_{\tilde{\mathcal{H}}}\nonumber\\
&&+\frac{1}{N^2}\sum_{\mathbf{k}_1\mathbf{k}_2}\left|
\tilde{y}'_{\mathbf{k}_1\mathbf{k}_2,\mathbf{k}-\mathbf{k}_1+\mathbf{k}_2}\right|^2
\langle f^\dagger_{\mathbf{k}-\mathbf{k}_1+\mathbf{k}_1}
f^{}_{\mathbf{k}-\mathbf{k}_1+\mathbf{k}_2}\rangle_{\tilde{\mathcal{H}}}\nonumber\\
&&\quad\times\langle c^\dagger_{\mathbf{k}_1}c^{}_{\mathbf{k}_1}\rangle_{\tilde{\mathcal{H}}}
\left(1-\langle c^\dagger_{\mathbf{k}_2}c^{}_{\mathbf{k}_2}\rangle_{\tilde{\mathcal{H}}}\right)\,,
\end{eqnarray}
\begin{eqnarray}\label{Eq:nfc}
{\mit{d}}^{}_{\mathbf{k}}&=&
\tilde{x}^{}_{\mathbf{k}}\tilde{x}'_{\mathbf{k}}
\langle f^\dagger_{\mathbf{k}}c^{}_{\mathbf{k}}\rangle_{\tilde{\mathcal{H}}}\nonumber\\
&-&\frac{1}{N^2}\sum_{\mathbf{k}_1\mathbf{k}_2}
\tilde{y}'_{\mathbf{k}_1\mathbf{k}_2,\mathbf{k}-\mathbf{k}_1+\mathbf{k}_2}
\tilde{y}^{}_{\mathbf{k}_1\mathbf{k},\mathbf{k}-\mathbf{k}_1+\mathbf{k}_2}
\langle c^\dagger_{\mathbf{k}_1}c^{}_{\mathbf{k}_1}\rangle_{\tilde{\mathcal{H}}}\nonumber\\
&&\times\langle f^\dagger_{\mathbf{k}_2}c^{}_{\mathbf{k}_2}\rangle_{\tilde{\mathcal{H}}}
\langle f^\dagger_{\mathbf{k}-\mathbf{k}_1+\mathbf{k}_2}
f^{}_{\mathbf{k}-\mathbf{k}_1+\mathbf{k}_2}\rangle_{\tilde{\mathcal{H}}}\,.
\end{eqnarray}
In the last step, one has to evaluate the expectation values
on the r.h.sides of Eqs.~\eqref{Eq:ncc}, \eqref{Eq:nff}, and \eqref{Eq:nfc}, which are
 formed with $\tilde{\mathcal H }$. Using the diagonal form of $\tilde{\mathcal H}$ in \eqref{HFanoRe}
one easily finds~\cite{PVN09}
\begin{align}
\label{ReNN1}
\langle c^\dagger_{\mathbf{k}}c^{}_{\mathbf{k}}\rangle_{\tilde{\mathcal H}}
&=u^2_{\mathbf k}f(E^c_{\mathbf k})+v^2_{\mathbf k}f(E^f_{\mathbf k})\,,\\
\langle f^\dagger_{\mathbf{k}}f^{}_{\mathbf{k}}\rangle_{\tilde{\mathcal H}}
&=v^2_{\mathbf k}f(E^c_{\mathbf k})+u^2_{\mathbf k}f(E^f_{\mathbf k})\,,
\label{ReNN}\\
\langle f^\dagger_{\mathbf{k}}c^{}_{\mathbf{k}}\rangle_{\tilde{\mathcal H}}
&=-[f(E^c_{\mathbf k})-f(E^f_{\mathbf k})]
\textrm{sgn}(\tilde{\varepsilon}^f_{\mathbf k}
-\tilde{\varepsilon}^c_{\mathbf k})
\frac{\tilde{\Delta}_{\mathbf k}}{W_{\mathbf k}}\,,
\label{ReNN2}
\end{align}
where $f(E^{\eta}_{\mathbf k})$ is the Fermi function.

\subsection{Spectral functions}

Let us consider the one-particle spectral function  for $c$ electrons
\begin{equation}
A^{c}({\mathbf k},\omega)=-\dfrac{1}{\pi}\textrm{Im}G^c({\mathbf k},\omega),
\end{equation}
where $G^c({\mathbf k},\omega)$ is the Fourier transform of the retarded Green function 
\begin{equation}
G^c({\mathbf k},\omega)=\langle\langle c^{}_{\mathbf{k}};c^{\dagger}_{\mathbf{k}} \rangle\rangle^{}(\omega+i0^{+}).
\end{equation}
Using again relation \eqref{erw}, the Green function can be rewritten
\begin{equation}
\langle\langle c^{}_{\mathbf{k}};c^{\dagger}_{\mathbf{k}} \rangle\rangle^{}(\omega)=
\langle\langle {c}^{}_{\mathbf{k}}(\lambda \rightarrow 0) ;
c^{\dagger}_{\mathbf{k}}(\lambda \rightarrow 0) \rangle\rangle^{}_{\tilde{\mathcal{H}}}(\omega)\,,
\end{equation}
where the expectation value on the r.h.side is again formed with 
$\tilde{\mathcal H}$.  Using  expression (\ref{cRe}) for $c^\dagger_{\mathbf k}(\lambda \rightarrow 0)$, 
we are immediately led the following result for the 
$c$-electron spectral function 
\begin{align}
\label{ImGcc}
A^{c}&(\mathbf{k},\omega)=|\tilde{x}^{}_{\mathbf{k}}|^2\left[u^2_{{\mathbf k}}\delta(\omega-E^c_{{\mathbf k}})+v^2_{{\mathbf k}}\delta(\omega-E^f_{{\mathbf k}})\right]\nonumber\\
+&\frac{1}{N^2}\sum_{{\mathbf k}_1{\mathbf k}_2}|\tilde{y}^{}_{\mathbf{k}_1\mathbf{k}\mathbf{k}_2}|^2\delta\left(\omega-(E^c_{\mathbf{k}_1}-E^f_{{\mathbf k}_1+{\mathbf k}_2-{\mathbf k}}+E^f_{\mathbf{k}_2})\right)\nonumber\\
&\times\Big[\langle c^\dagger_{\mathbf{k}_1}c^{}_{\mathbf{k}_1}\rangle_{\tilde{\mathcal{H}}}\left(\langle f^\dagger_{\mathbf{k}_2}f^{}_{\mathbf{k}_2}\rangle_{\tilde{\mathcal{H}}}-
\langle f^\dagger_{\mathbf{k}_1+\mathbf{k}_2-\mathbf{k}}f^{}_{\mathbf{k}_1+\mathbf{k}_2-\mathbf{k}}
\rangle_{\tilde{\mathcal{H}}}\right)\nonumber\\
&+\langle f^\dagger_{\mathbf{k}_1+\mathbf{k}_2-\mathbf{k}}f^{}_{\mathbf{k}_1+\mathbf{k}_2-\mathbf{k}}
\rangle_{\tilde{\mathcal{H}}}\left(1-\langle f^\dagger_{\mathbf{k}_2}f^{}_{\mathbf{k}_2}\rangle_{\tilde{\mathcal{H}}}\right)\Big]\,.
\end{align}
Note that we have restricted ourselves to the leading order in the EI order parameter.
In the same way, we can also calculate the spectral function $A^{f}(\mathbf{k},\omega)$
for the $f$-electrons. The final result reads
\begin{align}\label{ImGff}
A^{f}&(\mathbf{k},\omega)=|\tilde{x}'_{\mathbf{k}}|^2\left[v^2_{{\mathbf k}}\delta(\omega-E^c_{{\mathbf k}})+u^2_{{\mathbf k}}\delta(\omega-E^f_{{\mathbf k}})\right]\nonumber\\
&+\frac{1}{N^2}\sum_{{\mathbf k}_1{\mathbf k}_2}|\tilde{y}'_{\mathbf{k}_1\mathbf{k}_2,\mathbf{k}-\mathbf{k}_1+\mathbf{k}_2}|^2\nonumber\\
&\times\delta [\omega-(E^f_{{\mathbf k}-{\mathbf k}_1+{\mathbf k}_2}-E^c_{\mathbf{k}_2}+E^c_{\mathbf{k}_1})]\nonumber\\
&\times\Big[\langle c^\dagger_{\mathbf{k}_1}c^{}_{\mathbf{k}_1}\rangle_{\tilde{\mathcal{H}}}\left(1-\langle c^\dagger_{\mathbf{k}_2}c^{}_{\mathbf{k}_2}\rangle_{\tilde{\mathcal{H}}}\right)\nonumber\\
&\quad\;\;+\left(\langle c^\dagger_{\mathbf{k}_2}c^{}_{\mathbf{k}_2}\rangle_{\tilde{\mathcal{H}}}-
\langle c^\dagger_{\mathbf{k}_1}c^{}_{\mathbf{k}_1}\rangle_{\tilde{\mathcal{H}}}\right)\nonumber\\
&\qquad\qquad\times\left(1-\langle f^\dagger_{\mathbf{k}-\mathbf{k}_1+\mathbf{k}_2}f^{}_{\mathbf{k}-\mathbf{k}_1+\mathbf{k}_2}
\rangle_{\tilde{\mathcal{H}}}\right)\Big]\,.
\end{align}
\section{Numerical results and discussion}
We now will evaluate the analytical expressions of the PRM approach  
outlined so far. Of course, the set of  
equations~\eqref{Eq:ncc},~\eqref{Eq:nff}, and~\eqref{Eq:nfc}
has to be solved numerically. To this end, we choose some 
initial values for $\langle n^{c}_{\mathbf{k}}\rangle$, 
$\langle n^{f}_{\mathbf{k}}\rangle$, and 
$d^{}_{\mathbf{k}}$ 
(assuming $\langle c^\dagger_{\mathbf{k}}f^{}_{\mathbf{k}}\rangle=
\langle f^\dagger_{\mathbf{k}}c^{}_{\mathbf{k}}\rangle$),
and determine the renormalization of the Hamiltonian and all 
operators, by solving the differential 
equations~\eqref{ReEqs1}--\eqref{ReEqs3}, \eqref{ReEqsC}, 
and \eqref{ReEqsF}. Performing the limit $\lambda\rightarrow 0$, 
all model parameters will be renormalized. 
Then, using $\tilde{\mathcal{H}}$, the new expectation 
values [Eqs.~\eqref{ReNN1}-\eqref{ReNN2}] are calculated,  
and the renormalization 
process of the Hamiltonian is restarted. Convergence is assumed 
to be achieved if all quantities are determined with a relative 
error less than $10^{-5}$. The dynamical correlation 
functions~\eqref{ImGcc}--\eqref{ImGff} are evaluated using a 
Gaussian broadening in energy space of width 0.06. 
Because of the large number of differential equations that have 
to be solved, we confine ourselves, in what follows,  
to the investigation of the 1D case and limit 
the number of lattice sites ($k$-points) to $N=60$.
\subsection{Ground-state properties}
\subsubsection{Order parameter}
We begin by scanning the parameter space of the 1D EFKM, 
in order to detect an EI ground state. The $T=0$ quantum 
phase diagram of the 1D EFKM has been previously explored by the
CPMC technique~\cite{ZCG94}, after 
mapping the EFKM---rewritten in pseudospin variables---into a 
negative $U$ asymmetric Hubbard model with the Zeeman term replaced 
by a chemical potential~\cite{BGBL04}. Thereby, in terms of our original 
language, a transition from a mixed-valence regime to a nonmixed valence
regime was observed, which corresponds to the transition from 
the EI to a band insulator. In 1D, the EI phase is characterized by 
critical excitonic correlations. Surprisingly the topology of the 1D 
phase diagram is the same as for the 2D and 3D cases, which 
were studied by CPMC~\cite{BGBL04} (2D) and 
Hartree-Fock~\cite{Fa08,SC08} (2D, 3D) approaches. 
To benchmark the reliability of the PRM, we have compared the 1D PRM 
EI band-insulator transition points with those obtained by the rather unbiased 
CPMC method, and found excellent agreement. For example,  
we obtain 
$\varepsilon^f_{c,\text{PRM}}=-1.81\simeq \varepsilon^f_{c,\text{CPMC}}=-1.80$ 
for $t^f=-0.3$, $U=1$ (cf. Fig.~1 from Ref.~\onlinecite{BGBL04}).  
\begin{figure}[t]
    \begin{center}
      \includegraphics[angle = -0, width = 0.470\textwidth]{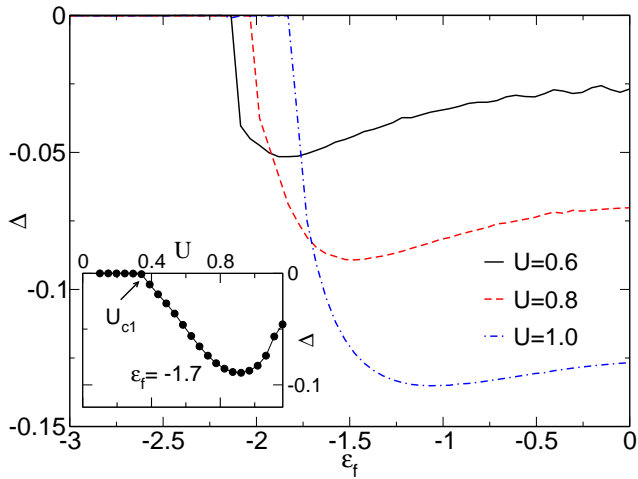}
    \end{center}
\caption{(Color online) EI order parameter, $\Delta$, in the 1D EFKM. 
The $f$-electron transfer-integral is fixed to be $t^f=-0.3$ 
(all energies are given in units of $t^c$). In what follows, 
we assume $\varepsilon^c=0$ without loss of generality.
The main panel gives $\Delta$ as a function of the position 
of the $f$-electron level, $\varepsilon^f$, for different 
values of $U$, while the inset shows the variation of 
$\Delta$ with $U$ at $\varepsilon^f=-1.7$. The calculations 
were performed at basically zero temperature, $T=10^{-3}$.}
\label{fig:DU}
\end{figure}

\begin{figure}[b]
    \begin{center}
      \includegraphics[angle = -0, width = 0.470\textwidth]{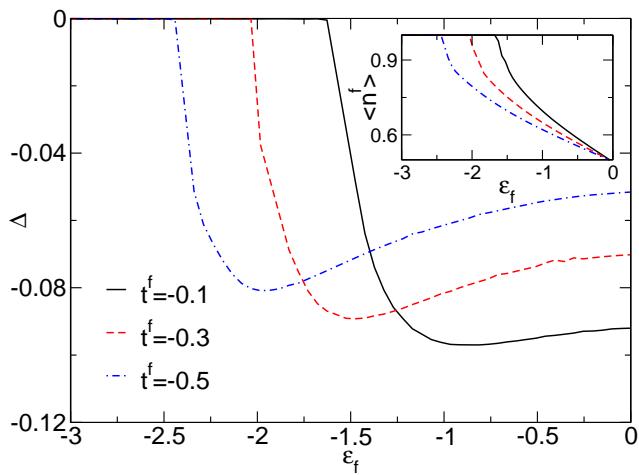}
    \end{center}
\caption{(Color online) EI order parameter in the 1D EFKM. The main panel 
gives $\Delta$ as a function of the position of the 
$f$-electron level for different $t^f$ at $U=0.8$. The inset 
shows the corresponding variation of the mean $f$-electron number.}
\label{fig:Df}
\end{figure}

Figures~\ref{fig:DU} and~\ref{fig:Df} show the onset of the EI phase 
for different values of $U$ ($t^f=-0.3$ fixed) and $t^f$
($U=0.8$ fixed), respectively, as the $f$-electron 
level is varied (see main panels). Obviously, the EI phase
emerges above a critical Coulomb attraction strength $U_{c1}$ 
(cf.~inset Fig.~\ref{fig:DU}), provided that $c$ and $f$ bands overlap. 
Moving up the $\varepsilon^f$ level, the top of the $f$ band  
reaches the bottom of the $c$ band at a critical value $\varepsilon^f_c$. 
Then some $f$ electrons can be transferred into $c$-band electrons
and exciton bound-states of $f$-band holes and $c$-band 
electrons may form if the Coulomb attraction is sufficiently strong.
Recent Hartree-Fock- and slave-boson-theory-based 
studies~\cite{SC08,IPBBF08,ZIBF10} yield a second, upper 
critical value of the Coulomb attraction $U_{c2}$, such that 
the EI phase is confined in between $U_{c1}$ and $U_{c2}$. 
The PRM data, produced up to now, will rather not confirm this 
controversial finding (see inset Fig.~\ref{fig:DU}). 
However in the large-$U$ limit, the numerical calculations are ill-conditioned 
and tedious, especially in 1D.  So the question whether  
$U_{c2}=\infty$ remains open.  That the appearance of the EI phase 
is intimately connected with the build-up of $f$-$c$ electron coherence 
and a non-integral $f$-electron valence is demonstrated by the 
inset of Fig.~\ref{fig:Df}, depicting $\langle n^f\rangle$ 
(cf. also the discussion of Fig.~\ref{fig:Nk} below).

The physical picture developed so far does not change
if the $f$-electron band approaches the $c$-electron band from above.
That is our results are one and the same changing the sign of
$\varepsilon^f$. We note that in a 
rather small region around $\varepsilon^f=0$ (symmetric band
case), a charge-density-wave state is assumed to be the true ground state,
i.e., the EI phase becomes metastable. Focussing on the characterization 
of the EI phase we will not address this issue here.

\subsubsection{Band renormalization}
\begin{figure}[t]
    \begin{center}
      \includegraphics[angle = -0, width = 0.470\textwidth]{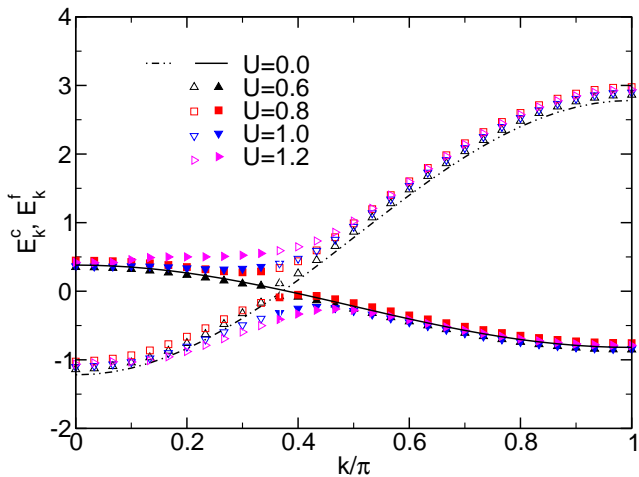}
    \end{center}
\caption{(Color online) Renormalized quasiparticle 
band dispersion of $c$ electrons 
($E^c_k$, open symbols) and $f$ electrons ($E^f_k$, filled symbols) 
in the EI phase of the 1D EFKM for different values of $U$. 
Again the `bare' band structure is parameterized by  
$\varepsilon^c=0$, $\varepsilon^f=-1.0$, $t^f=-0.3$ (dot-dashed and solid lines). 
Note that the scale of the ordinate is shifted in order
to fix the Fermi energy at zero energy.}
\label{fig:Ek}
\end{figure}

Next we investigate the renormalization of the 1D
band structure.  Figure~\ref{fig:Ek} displays the 
$k$-dependence of the quasiparticle energies $E_k^\eta$
in the EI phase, where open and filled symbols correspond
to bands having predominantly $c$- and $f$-electron character, 
respectively. There are two features of importance. First,
the band structure is clearly gapful in the EI phase. 
Hence, for the half-filled band case, the system is insulating. 
The gap originates from $c$-$f$ electron hybridization, 
induced by the attractive Coulomb interaction.  
Second, the lower and upper quasiparticle bands are narrowed 
as a result of the electronic correlations. 
While the maximum of the lower (at $T=0$ completely occupied) 
band is displaced to larger $k$-values as $U$ increases,  
the minimum of the upper (at $T=0$ empty) band moves to smaller $k$, 
accompanied by a flattening of the `$f$-band' near $k=0$.
At very large $U$ (not depicted), the gap is kept (but different in nature),
because of the extreme Hartree shift, leading to a $c$-$f$ band-splitting. 
Except of contributions from the band narrowing, 
the Hartree shift to the quasiparticle 
energies is roughly given by the sum of  
${\varepsilon}^c - {\varepsilon}^f$ and 
$U(\langle n^f \rangle - \langle n^c \rangle  )$
(compare Eqs.~(\ref{hartreeshift}), (\ref{qpe_c}), (\ref{qpe_f})).

\subsubsection{Momentum distribution function}
\begin{figure}[t]
    \begin{center}\vspace*{0.2cm}
      \includegraphics[angle = -90, 
width = 0.470\textwidth]{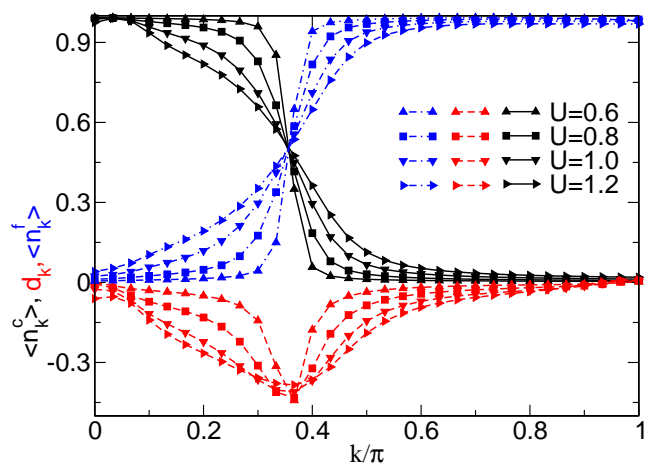}
    \end{center}
\caption{(Color online) Momentum distribution functions 
$\langle n^c_k\rangle=\langle c^\dagger_{{k}}c^{}_{{k}}\rangle$ 
(black solid lines) and 
$\langle n^f_k\rangle=\langle f^\dagger_{{k}}f^{}_{{k}}\rangle$ 
(blue dash-dotted lines) for the same model parameters as used
in Fig.~\ref{fig:Ek}. The red dashed lines show the corresponding
`order parameter' functions $d^{}_{k}$, 
see Eq.~\eqref{Delta}.}
\label{fig:Nk}
\end{figure}

Another quantity of interest is the occupation number of fermionic
states carrying momentum $k$. For free fermions, at $T=0$,
all states up to the Fermi energy, $E_F$, are occupied, so that the
momentum distribution function, $n(k)=\langle n_k\rangle$,
has a discontinuity at the corresponding Fermi momentum, $k_F$,
where $n(k)$ jumps from one to zero . In an interacting Fermi liquid 
there is still a discontinuity, but the jump is less than one.
In 1D, normally Luttinger-liquid behavior emerges, with an essential 
power-law singularity at $k_F$. For the insulating state, however, $n(k)$ 
is given by a smooth curve. This holds, e.g., for the charge-density-wave
ground states of 1D $t-V$ and Holstein-type models~\cite{EGN05},
and should also be valid for the EI phase in the 1D EFKM. Indeed,
the momentum distribution functions of $c$ and $f$ electrons, $n^\eta(k)$,
depicted in Fig.~\ref{fig:Nk} confirm this picture. We see that
$n^f(k)$ ($n^c(k)$) monotonously increases (decreases) as $k$ varies
from $k=0$ to $k=\pi$. As expected, the drop (upturn) near `$k_F$' 
softens at larger interactions strengths. 

We have also included in Fig.~\ref{fig:Nk} the variation of 
${\mit{d}}^{}_{k}=\langle f^\dagger_{{k}}c^{}_{{k}}\rangle$. 
This expectation value enters into the equation for the 
order parameter~\eqref{Delta}. In some sense, it can be taken as 
a measure of the range in $k$-space, where $c$-electrons and $f$-holes
are involved in the exciton formation and condensation process.   
Having a $U$-driven BCS-BEC crossover scenario in the EI phase 
of the EFKM in mind~\cite{BF06,IPBBF08}, the broadening of the  
distribution of ${\mit{d}}^{}_{k}$ with increasing $U$ might indicate  
the condensation of a more local two-body BEC-like bound-state out of a  
BCS-like Cooper-pair state. 
Note that in the BEC-like state the
Fermi surface plays no role.   

\subsection{Spectral properties}
In this section, we present first results for the photoemission (PE) spectra
in the EI phase of the EFKM. The calculated single-particle spectral 
functions, associated with the emission (PE) or injection (inverse PE) 
of an electron with wave vector $k$, directly measure the occupied and 
unoccupied densities of single-particle states, and therefore are well
suited to investigate pairing gaps as well. Note that the ARPES spectral functions
\begin{eqnarray}
A^\eta_{\text{ARPES}}({\mathbf k}, \omega) &=& \frac{1}{2\pi} \int_{-\infty}^{+ \infty}
\langle \eta_{\mathbf k}^\dagger (t) \eta_{\mathbf k} \rangle \, e^{- i \omega t} dt   \nonumber\\
&=&    \frac{1}{1+ e^{ \beta \omega}} A^\eta ({\mathbf k}, \omega) 
\end{eqnarray}
fulfill at $T=0$ the frequency sum rule
 \begin{equation}
\label{sumrule}
 \int_{-\infty}^0 A^\eta({\mathbf k}, \omega) 
d\omega =   \langle n_{\mathbf k}^\eta \rangle 
\end{equation}
where $ \langle n_{\mathbf k}^\eta \rangle$ is given by Eqs.~\eqref{Eq:ncc} and \eqref{Eq:nff}
(cf. Fig.~\ref{fig:Nk}).

\begin{figure*}[t]
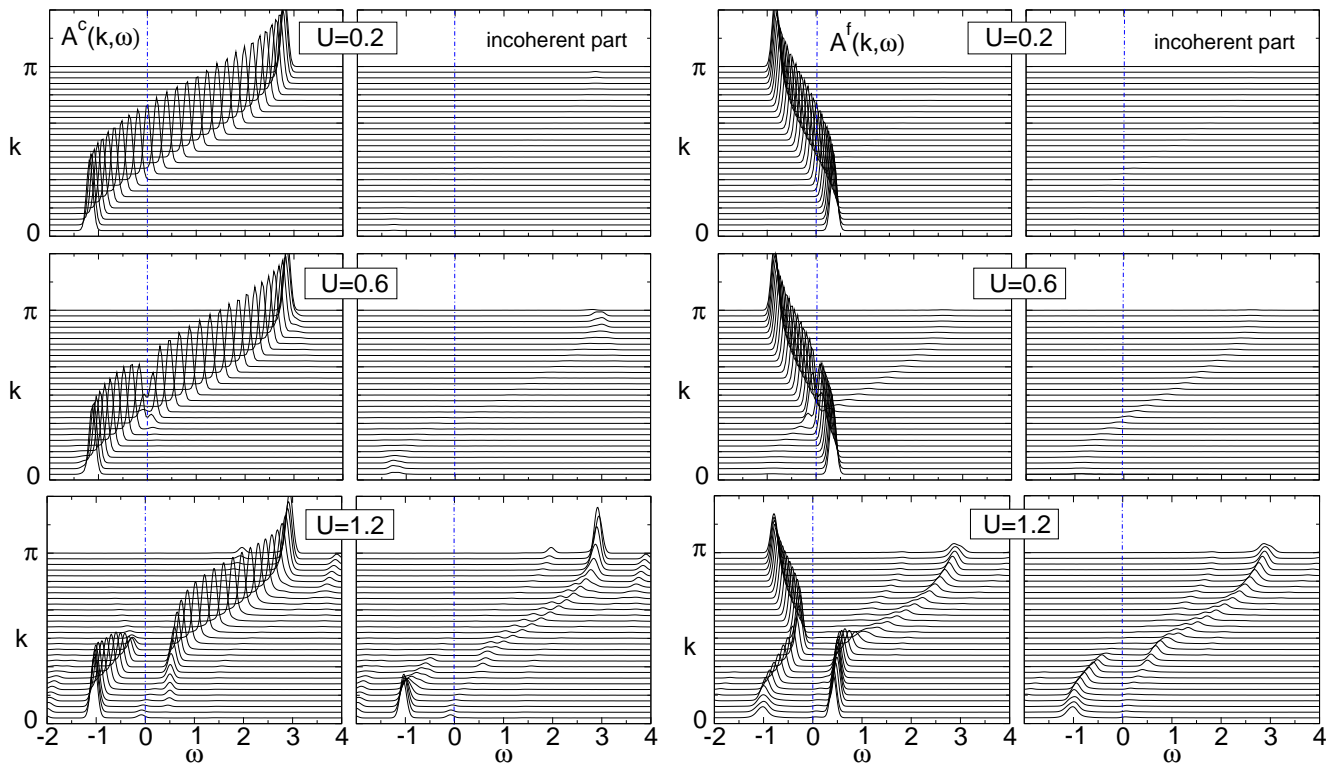

    \begin{center}
      \includegraphics[angle = -0, width = 0.48\textwidth]{fig6a.eps}
\hspace*{0.18cm}\includegraphics[angle = -0, width = 0.48\textwidth]{fig6b.eps}
\vspace{0.05cm}\\
      \includegraphics[angle = -0, width = 0.48\textwidth]{fig6c.eps}
\hspace*{0.18cm}\includegraphics[angle = -0, width = 0.48\textwidth]{fig6d.eps}
\vspace{0.05cm}\\          
\hspace*{0.06cm}\includegraphics[angle = -0, width = 0.48\textwidth]{fig6e.eps}
\hspace*{0.18cm}\includegraphics[angle = -0, width = 0.48\textwidth]{fig6f.eps}
    \end{center}
\caption{(Color online) Wave-number resolved photoemission spectra of the 1D 
half-filled EFKM. The $c$- (left-hand panels) and $f$-electron 
(right-hand panels) single-particle spectral functions
were calculated  for several characteristic values of $U$
at $\varepsilon^f=-1.0$, $t^f=-0.3$, $T=10^{-3}$. 
The left panels show in each case the total spectra, $A^{\eta}(k,\omega)$,
whereas the right panels give the `incoherent' contributions only 
(second term in Eqs.~\eqref{ImGcc},~\eqref{ImGff}). 
The vertical dot-dashed lines mark the chemical potential.}  
\label{fig:Awcfef10}
\end{figure*}

Figure~\ref{fig:Awcfef10} displays the zero-temperature, wave-vector 
and energy resolved single-particle spectral functions, 
$A^\eta(k,\omega)$ [see Eqs.~\eqref{ImGcc}, \eqref{ImGff}],
for a bare band structure parameterized by $\varepsilon^f=-1$
($\varepsilon^c=0$), $t^f=-0.3$ ($t^c=1$).
For weak Coulomb attraction, $U=0.2$ (upper panels) 
we are still in the semi-metallic phase, and consistently
$A^c(k,\omega)$ and $A^f(k,\omega)$ follow the nearly unrenormalized
$c$- and $f$-band dispersions, respectively. Concomitantly, 
we find a more or less uniform distribution of the spectral 
weight and negligible incoherent contributions (see right-hand panels
for $A^c(k,\omega)$ and $A^f(k,\omega)$). When entering the
EI phase by increasing $U$ to $U=0.6$, a gap feature develops at the
Fermi energy (Fermi momentum), but away from that the spectra 
still show the main characteristics of the semi-metallic state 
(cf. both middle panels). At a still larger value $U=1.2$, the gap broadens.
Most notably, however, is a significant redistribution 
of the spectral weight from the coherent to the incoherent part
of $A^\eta(k,\omega)$, with pronounced absorption maxima
at $k=0,\,\pi$. Of particular importance is a considerable admixture of 
$c$-electron contributions to the $f$-electron spectrum in an interval between 
$k=0$ and $k \simeq k_F$. The resulting double peak structure around the 
Fermi level can be considered as an almost $k$-independent bound object of 
$c$ electrons and $f$ holes.   

\begin{figure*}[t]
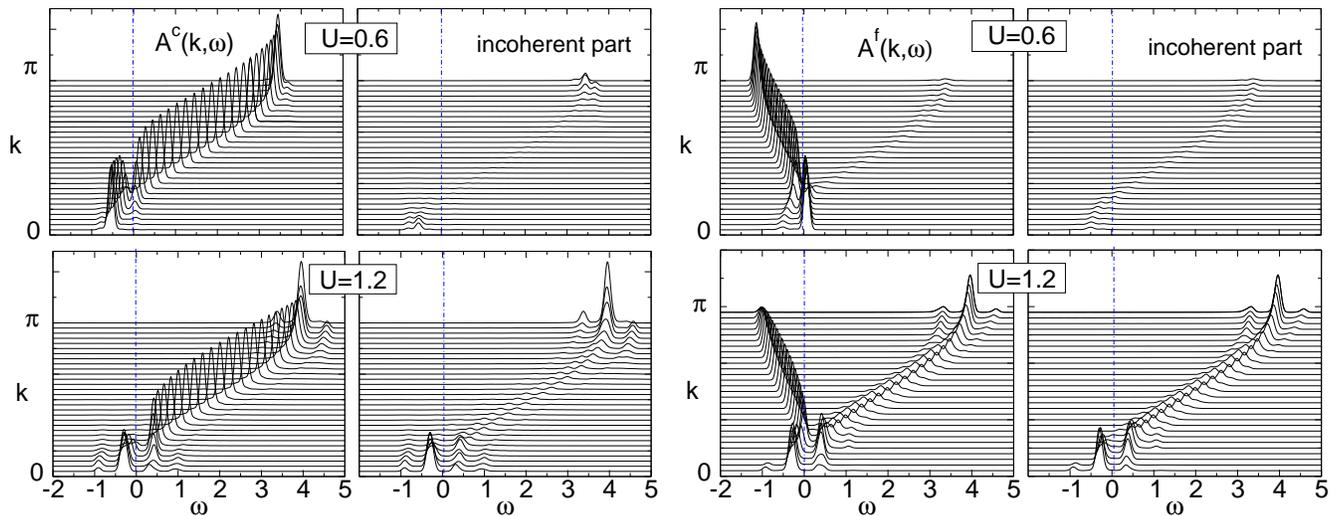

    \begin{center}
      \includegraphics[angle = -0, width = 0.48\textwidth]{fig7a.eps}
\hspace*{0.2cm}\includegraphics[angle = -0, width = 0.48\textwidth]{fig7b.eps}\\ 
 \hspace*{0.06cm}     \includegraphics[angle = -0, width = 0.48\textwidth]{fig7c.eps}
\hspace*{0.2cm}\includegraphics[angle = -0, width = 0.48\textwidth]{fig7d.eps}
    \end{center}
\caption{(Color online) Single-particle spectral functions for 
$c$ (left-hand panels) and $f$ electrons 
(right-hand panels) in 1D half-filled EFKM 
with $\varepsilon^f=-1.7$, $t^f=-0.3$, $T=10^{-3}$. 
As in Fig.~\ref{fig:Awcfef10}, the total spectra are contrasted 
to the `incoherent' contributions.}  
\label{fig:Awcfef17}
\end{figure*}

If we change the location of the $f$-band by 
lowering the position of the $f$-electron level, 
 we can achieve that the band 
structure becomes gapful due to the Hartree 
shift~\eqref{hartreeshift}, even for moderate values of $U$
(assuming $\Delta\equiv 0$). For example, at $\varepsilon^f=-1.7$ and 
$t^f=-0.3$, a $c$-$f$ band-splitting 
(positive Hartree gap $\Delta_H$, cf.~Fig.~\ref{fig:PD}) 
occurs already at $U_{H}\simeq 0.94$.   
Such a situation comes closer to the Tm[Se,Te] system~\cite{NW90}.  
Figure~\ref{fig:Awcfef17} shows the PE spectra calculated
for these parameter values. Compared to the case 
$\varepsilon^f=-1$, for $U=0.6$ the gap is clearly a bit more 
shaped, but the main features of the spectrum 
stay the same. Here we are in the BCS-regime, where pairing 
fluctuations are expected to be small. For $U=1.2> U_H$, 
we enter the BEC-regime, where preformed pairs acquire
quantum coherence (many-body character) during the condensation 
process. Since $\Delta \neq 0$ the gap persists. The distinct  
incoherent contributions, showing up in the $A^\eta(k,\omega)$ spectra 
at high energies, are related to the dissociation of 
two-particle bound-states (excitons). 

\begin{figure*}[t]
\includegraphics[angle = -90,
width =0.485\textwidth]{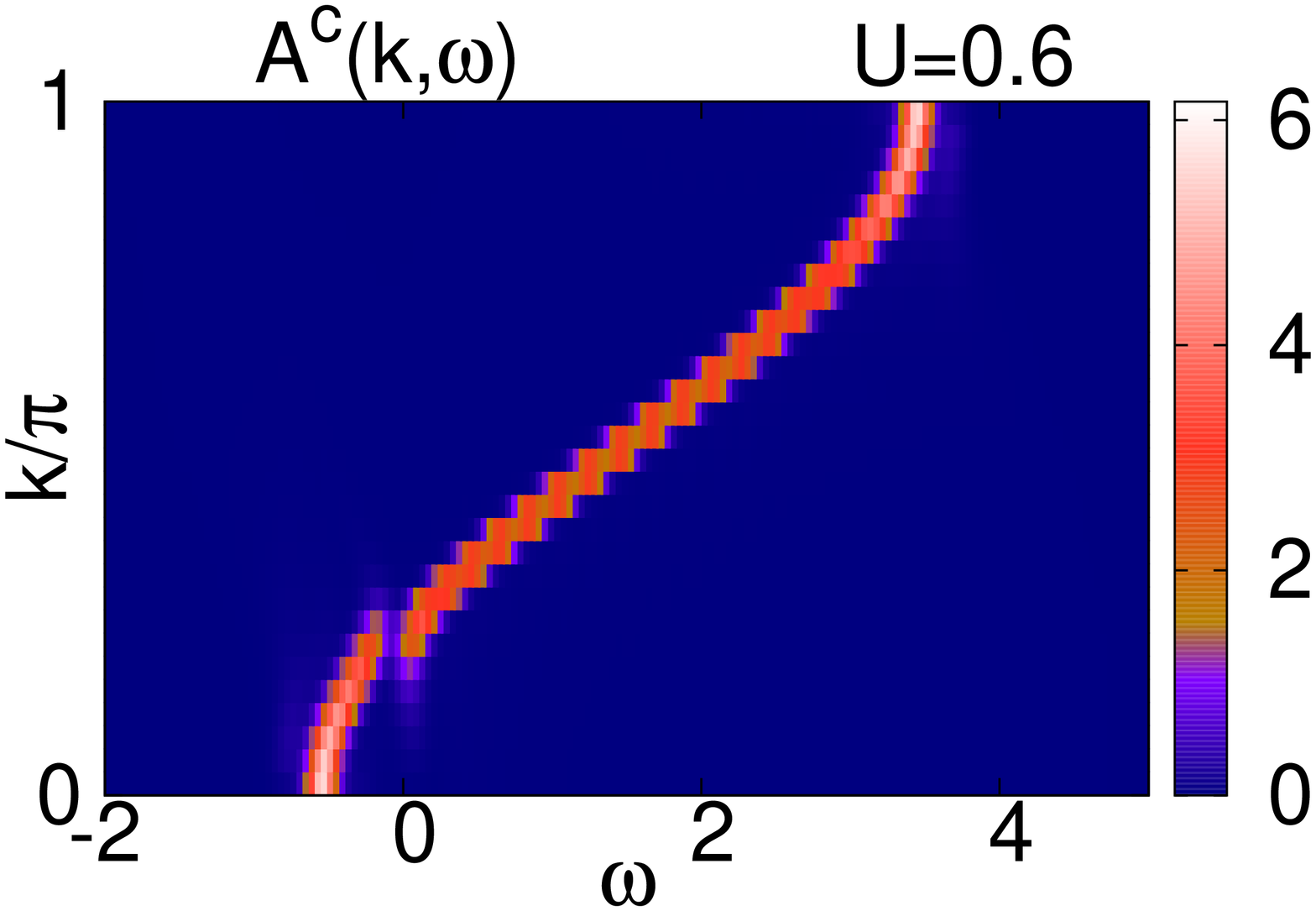}
\includegraphics[angle = -90,
width = 0.485\textwidth]{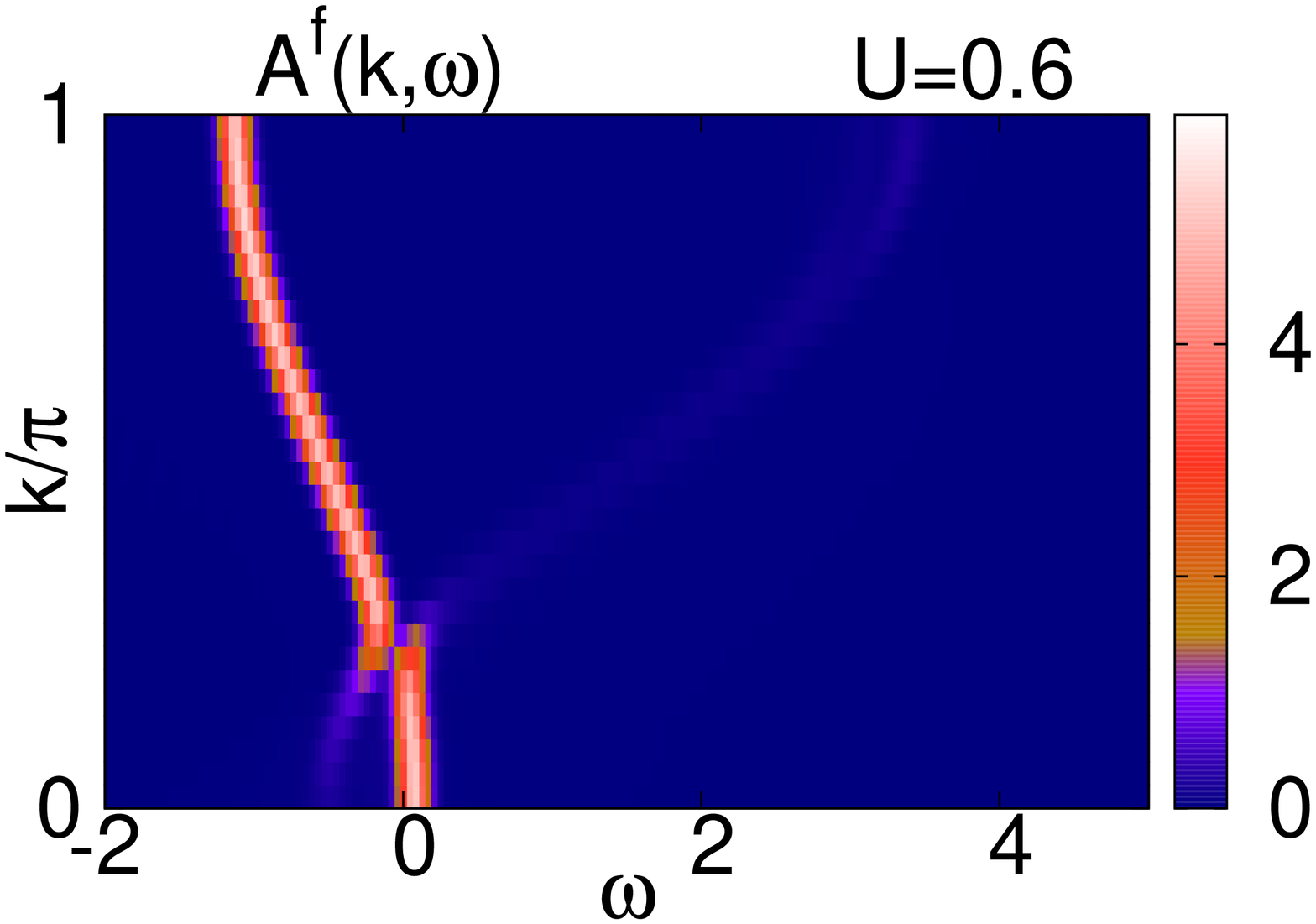}\\
\includegraphics[angle = -90,
width = 0.485\textwidth]{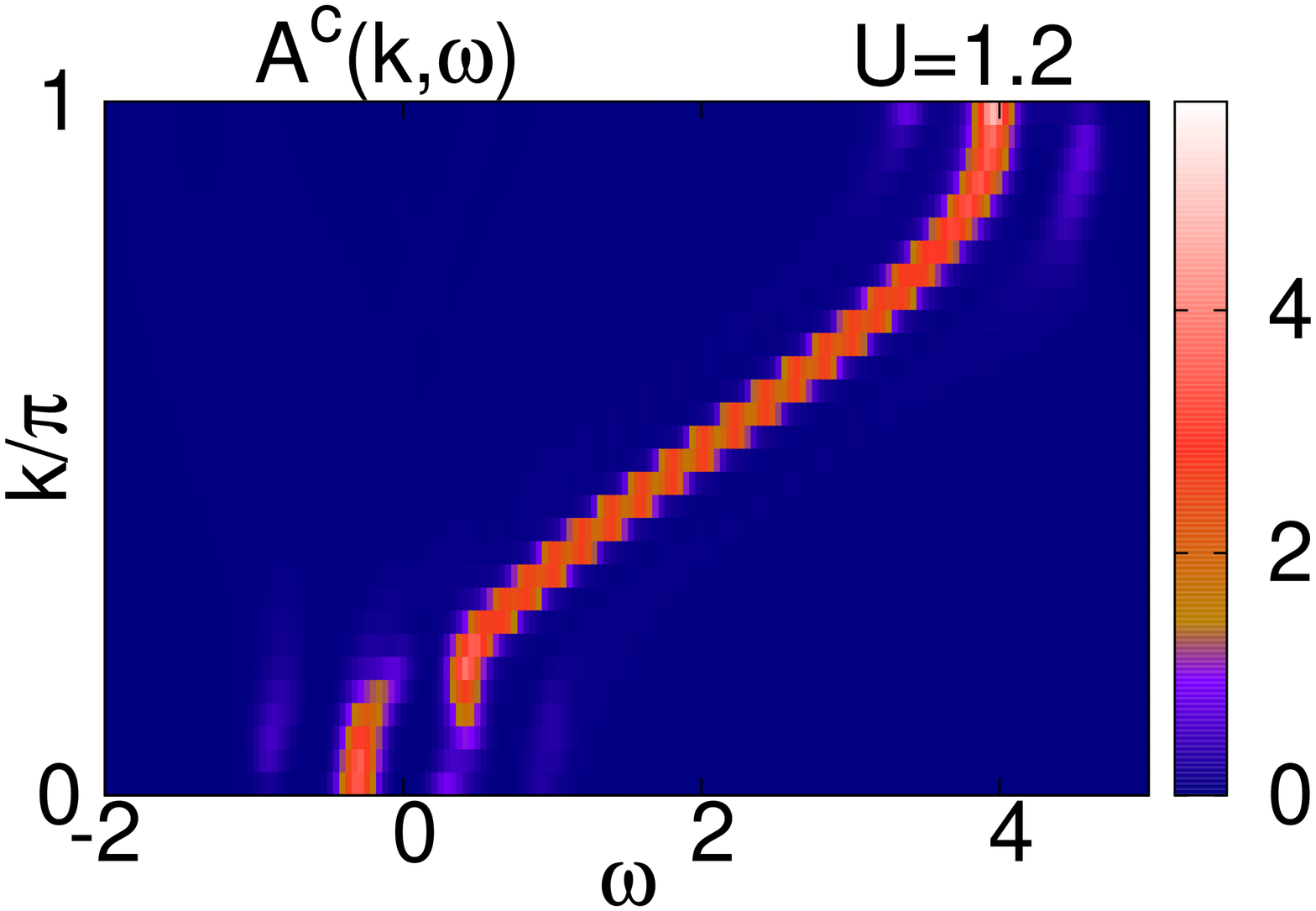}
\includegraphics[angle = -90,
width = 0.485\textwidth]{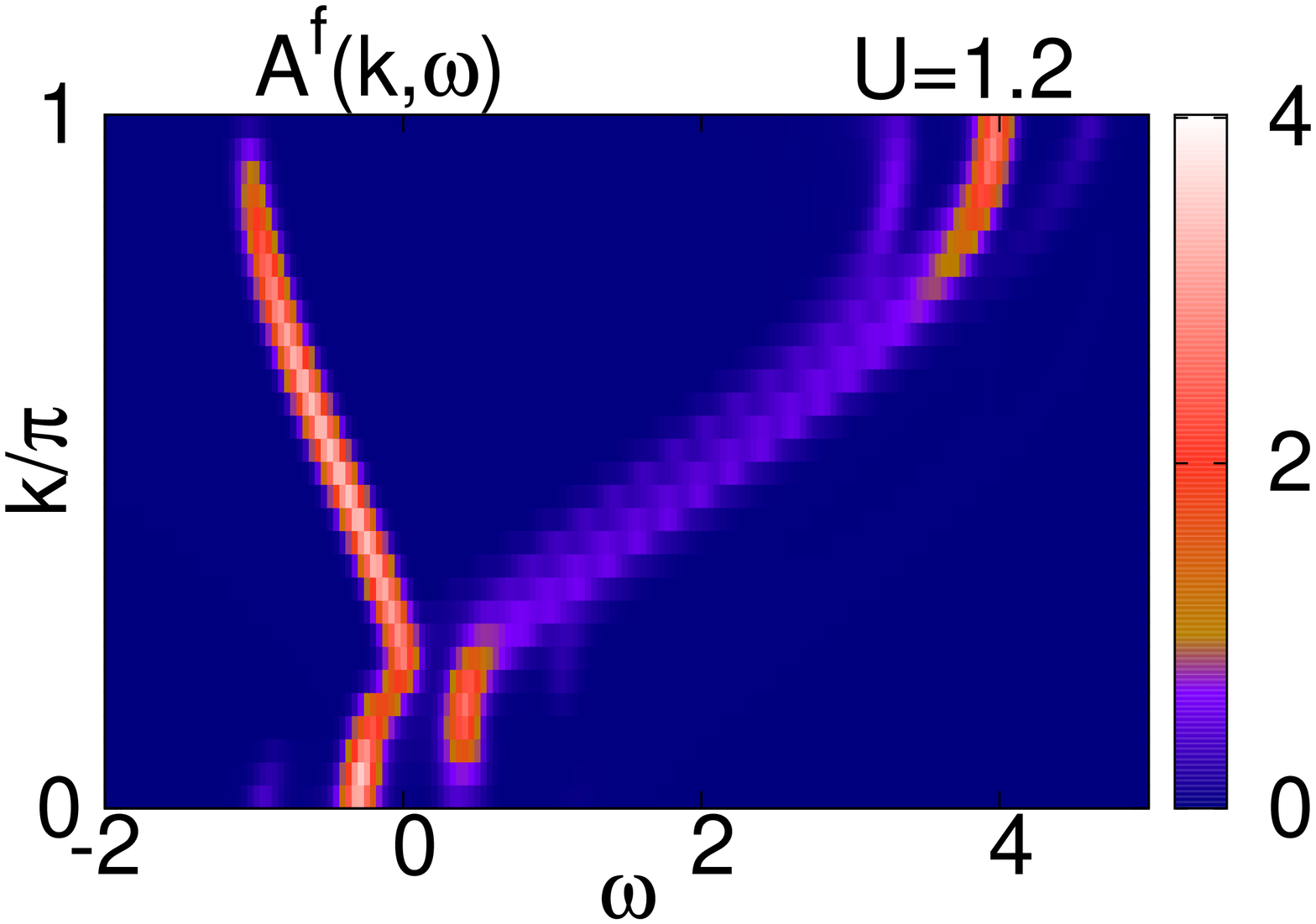}
\caption{(Color online) Intensity plots of the $c$- (left-hand panels) and $f$- 
(right-hand panels) electron spectral functions 
$A^{\eta}(k,\omega)$, presented in Fig.~\ref{fig:Awcfef17}, for
$\varepsilon^f=-1.7$, $t^f=-0.3$ at $U=0.6$ (upper panels) and $U=1.2$ 
(lower panels).}
\label{fig:InAwcfef17}  
\end{figure*}
To visualize more clearly the spectral weight and line-shape
of the various absorption signals, we provide in 
Fig.~\ref{fig:InAwcfef17} a color-map intensity-plot of the 
single-particle spectral functions depicted in Fig.~\ref{fig:Awcfef17}. 
Particularly from the $A^f(k,\omega)$ data, 
it appears that the BCS-BEC crossover is again marked by 
a notable admixture of $c$-electron-like contributions leading to new
peak-line structures in the spectrum (cf.~right-hand panels of Fig.~\ref{fig:InAwcfef17}).

\section{Conclusions}
To summarize, we adapted a continuous version of 
the projective renormalization method
to the extended Falicov-Kimball model to examine the possible
existence of a collective excitonic insulator phase at the 
semimetal-semiconductor transition. Thereby the PRM approach 
allows us to derive analytical expressions for the EI order parameter
and various other physical quantities characterizing the ground-state
and spectral properties of the model. The self-consistent evaluation of 
the renormalization equations yields a stable EI solution for the 
one-dimensional EFKM, at half-filling and zero temperature. It
therefore confirms previous, straight numerical data by constrained
path integral Monte Carlo. In particular, the phase boundary between
the excitonic and band insulator agrees even quantitatively with
the CPMC results. Thus, increasing the Coulomb attraction 
between $c$-band electrons and $f$-band holes, the appearance of 
a semimetal-EI transition seems 
to be settled for the 1D EFKM. 

Moreover, we present novel results for the single-particle  
$c$- and $f$-electron spectral function of the EFKM. The calculated
photoemission spectra are in evidence of a BCS-BEC crossover of 
the excitonic condensate, triggered by the Coulomb interaction.
Thereby the character of the electron-hole pairs changes from the
many-body bound state associated with the Cooper-type instability 
(weak-to-intermediate coupling BCS-side) to the two-body 
(tightly-bound exciton) bound-state (intermediate-to-strong coupling
BEC side), where Fermi surface effects are negligible. 
Hallmark of the BCS-BEC crossover in the quasiparticle spectra
is a substantial spectral weight transfer from the coherent to the 
incoherent part of the spectrum. A more thorough analysis of the
pairing fluctuations, which are expected to be strongly 
enhanced in the BCS-BEC transition region, e.g. by calculating
the dynamical pair-susceptibilities within the PRM approach, 
would be a worthwhile goal of forthcoming studies.

\begin{acknowledgements}
The authors thank F. X. Bronold, B. Bucher, D. Ihle, C. Monney, 
 P. Wachter, and B. Zenker for valuable discussions. 
This work was supported by the DFG through the research program 
SFB 652, B5.
\end{acknowledgements}


\end{document}